\documentclass[nofootinbib,preprintnumbers,amsmath,amssymb,twocolumn]{revtex4}

\usepackage{xcolor}

\newcommand{\ket}[1]{|#1\rangle}
\newcommand{\bra}[1]{\langle#1|}
\newcommand{\be}{\begin{equation}}
\newcommand{\ee}{\end{equation}}
\newcommand{\bea}{\begin{eqnarray}}
\newcommand{\eea}{\end{eqnarray}}

\newcommand{\tmpcolor}[1]{} 
\newcommand{\cor}[1]{{#1}} 
\newcommand{\alice}{{Alice}} 
\newcommand{\bob}{{Bob}}

\begin{document}

\title{2022 Nobel Prize in Physics and the End of Mechanistic Materialism}
\author{Igor Salom}
\affiliation{Institute of Physics, Belgrade \\ University, Pregrevica 118, Zemun, Serbia }

\begin{abstract}
The ideas and results that are in the background of the 2022 Nobel Prize in physics had an immense impact on our understanding of reality. Therefore, it is crucial that these implications reach also the general public, not only the scientists in the related fields of quantum mechanics. The purpose of this review is to attempt to elucidate these revolutionary changes in our worldview that were eventually acknowledged also by the Nobel's committee, and to do it with very few references to mathematical details (which could be even ignored without undermining the take-away essence of the text).

We first look into the foundational disputes between Einstein and Bohr about the nature of quantum mechanics, which culminated in the so-called EPR paradox -- the main impetus for all the research that would ensue in this context. Next, we try to explain the statement of the famous Bell's theorem -- the theorem that relocated the Einstain-Bohr discussions from the realm of philosophy and metaphysics to hard-core physics verifiable by experiments (we also give a brief derivation of the theorem's proof). Then we overview the experimental work of last year's laureates, which had the final say about who was right in the debate. The outcome of these experiments forced us to profoundly revise our understanding of the universe. Finally, we discuss in more detail the implications of such outcomes, and what are the possible ways that our worldviews can be modified to account for the experimental facts. As we will see, the standard mechanist picture of the universe is no longer a viable option, and can be never again. Nowadays, we know this with certainty unusual for physics, that only a strict mathematical theorem could provide.
\end{abstract}

\maketitle

\section{Introduction}

The 2022 Physics Nobel Prize was not quite like any other. While the Nobel prizes in physics are always of interest to the physics community, by a rule, they are merely a matter of curiosity for the general public. However, \cor{2022} Nobel award should pertain to all of us, irrespective of the profession, and remind us that it's been a time to rethink our basic worldviews. Indeed, the Nobel committee announcement in early October 2022 has elicited some bold titles across journals and news agencies worldwide, even among those not aimed at scientific circles. For example, the business newspaper Financial Times, on October 11, came up with an article titled ``What quantum physics tells us about reality'', while the nonprofit news website ``The Wire'' had a piece titled ``In the 2022 Physics Nobel Prize, a Test of Your Grip on Reality''. Scientific American, in its October issue, brought up the article ``The Universe Is Not Locally Real, and the Physics Nobel Prize Winners Proved It''. The deep philosophical impact of the 2022 Nobel Prize was particularly nicely emphasized in the title that last October showed up in ``The Conversation'' media network: ``How philosophy turned into physics – and reality turned into information''. As we see, the word ``reality'' is recurrent in these titles, and few things should be of more pervasive importance than the calls to reconsider our notion of the fundamental reality, especially when these calls (essentially) originate from the Royal Swedish Academy of Sciences.

So, in what sense do the accomplishments of \cor{2022} laureates Alain Aspect, John F. Clauser, and Anton Zeilinger challenge our understanding of reality? The Nobel committee summarizes that they \cor{were} awarded ``for experiments with entangled photons, establishing the violation of Bell inequalities and pioneering quantum information science'', but it takes some time and effort to unwrap this ostensibly simple statement. To do so, we must start with the achievements of some people who could not appear at the last year's Nobel ceremony.

\section{Einstein and Bohr}

\subsection{The debates}

The first act of the deep philosophical drama that had its (scientific) resolution in the last year's Nobel prize begins in the first half of the XX century, in years between the two world wars, when the ``new physics'' called Quantum Mechanics has begun to assume its mathematically well-defined, more or less present-day form. While there essentially existed a global consensus that the claims of the new theory seemed preposterous, there was a huge disagreement to what extent such a state of affairs reflected the true properties of nature and to what extent only of the present state of (possibly highly incomplete) scientific understanding of that nature. The opinions ranged from that of the camp of (some of) the ``founding fathers of quantum mechanics'', to whom belonged likes of Niels Bohr, Werner Heisenberg, and Wolfgang Pauli, all the way to staunch criticizers of quantum ideas among whom was no lesser figure than the Albert Einstein himself. While the former insisted that quantum mechanics provides an accurate and complete description of reality (and if we do not like it, it is nature to blame), the latter were convinced that quantum physics is at best a temporary and approximate model of some more ``reasonable'' underlying physics that we are yet to discover.

The new physics seemed to ignore all cornerstones of the previously known exact science. {\tmpcolor{purple}One of the more obvious departures from the previous scientific reasoning was the sudden appearance of randomness at the most fundamental level of the new theory -- prompting Einstein to famously protest with ``God does not play dice''. Namely, we must recall that one of the main tenets of all the prior physics, including then novel theory of relativity, was that there can be nothing inherently random in nature -- everything evolves according to strict mathematical laws that precisely \emph{determine} the future of any physical system, leaving no room for chance. Any randomness in the events that we might observe had to be a consequence of our imprecise knowledge either of the present state of the universe, or of the laws of nature, or a consequence of our technical inability to calculate the future from these basic laws. Yet, for a hypothetical being (so-called Laplace's daemon) who knows precise positions and momenta of every atom in the universe, and has both the complete knowledge of physical laws and sufficient computational capabilities -- all the future (and past also) was fixed and known. Such a view was essentially uncontested in science -- until this new quantum theory. Famously, in those days, Max Born put forward his ``probabilistic'' interpretation of the wavefunction (the wavefunction was the main mathematical object of the new physics, describing all properties of a system), according to which we often could do no more but to predict only the probabilities of experimental outcomes -- and not due to any lack of technological abilities, but in principle. At face value, the quantum theory suggested that physical laws simply do not fully determine the future of systems, but only provide probabilities for the possible events. This seemed outrageous to many, including Einstein. But, it was gradually becoming clear that much more is at stake than merely this \emph{determinism} of classical physics.}  

The novel ``uncertainty relations'', produced by Heisenberg \cor{-- stating that some  mutually} ``incompatible'' properties such as position and momentum cannot be simultaneously known with arbitrary precision -- seemed to be inevitably related to the proposed indeterminism. These relations provided an intuitive rationale for why we cannot reduce uncertainties in our knowledge of physical systems and again recover determinism. Intuitively, due to them was impossible to ``track down'' all details of the present state of the system, and thus it was also impossible to precisely predict its future state - allowing only for making experimental predictions of statistical type. At least, this was intuition based on our classical understanding of systems and their properties - which, of course, implied that these properties are well defined per se, in spite that our knowledge of them might be quite vague. However, even here the positions of the two camps started to diverge.

Indeed, it was not clear how to interpret the uncertainty relations themselves. It was one thing to accept that we are incapable to determine certain physical quantities with arbitrary precision, even to accept that the laws of physics happen to ``conspire'' in such a way that this turns out to be impossible in principle, that is, irrespective of our technological abilities and experimental skills. In particular, it seemed reasonable that the very act of measurement, which must involve some interaction with the object being measured, inescapably disturbs that object in an unpredictable fashion that leads to uncertainties. Einstein seemed quite ready to accept such a possibility, even if it would, in practice, preclude any prospect of determining precise properties of the systems and forever confine our experimental predictions to the mere computation of probabilities. We are imperfect creatures, and thus it is not inconceivable that our capacities, including those to gather (precise) data about the universe, are strictly limited. In this case, it is our imperfection and our problem. Universe (and systems themselves) certainly ``know'' their properties, and underlying objective reality must be well-defined, solid real. The uncertainty relations were in this view technical or epistemological limitations and had nothing to do with the true ontology, which remains to be described by a deeper, more complete theory than quantum mechanics. Consequently, Einstein maintained that deep down ``God does not play dice with the universe'', no matter how that might look to us. Even Heisenberg himself seemed to initially share this view of his own relations - as can be inferred from his analysis of what is nowadays known as Hesenberg's microscope \cite{HeisenbergMicroscope} (the position he soon after changed, under Bohr's influence \cite{HeisnbergAcceptingComplementarity}).

Yet, it was an entirely different thing to assume that the uncertainty relations do not merely tell about our inability to reveal the true properties of nature, but that they actually reflect the vagueness of these properties themselves, per se. That the problem is not that when the momentum of a particle is precisely known and defined, we \emph{cannot know} the position of this particle, but that this particle position in a certain sense no longer ``exists'', i.e.\ that it is not a well-defined property, of itself. (Alternatively, following the mathematical formalism, we might then say that this property becomes blurred into a \cor{so-called} superposition of various possibilities.) In other words, that neither the universe nor the particle itself really ``knows'' this location. And this was gradually becoming the position of the Bohr's camp. Bohr in particular saw this via his principle of complementarity, the precursor of the contemporary notion of contextuality in quantum physics. Knowledge of one of the two ``complementary'' properties (i.e.\ of those ``incompatible''\footnote{Mathematically, operators corresponding to such properties do not commute.} in the context of uncertainty relations) automatically renders the other one ill-defined: precise measurement of momentum does not simply make the position of the particle unknown or unpredictable (e.g. as a consequence of mechanically disturbing the particle) - it makes it fully meaningless and undefinable. In his view, the properties themselves were inseparable from the experimental \emph{context} that reveals them: the setup needed to measure the momentum crucially differs from the one required to find out the position - these two properties cannot be observed or measured simultaneously. Thus it was also wrong to assume that the position exists in the context where the momentum was (precisely) measured, and vice versa.

But such a view was inevitably leading further down the rabbit hole. It was obviously altering our \cor{basic} views about \emph{reality}, to cite \cite{BohrReality}: ``In the traditional view, it is assumed that there exists a reality in space-time and that this reality is a given thing, all of whose aspects can be viewed or articulated at any given moment. Bohr was the first to point out that quantum mechanics called this traditional outlook into question.'' Moreover, if the position, or momentum (and this also holds for basically any other property) is not always well-defined, when is it that it becomes well-defined? Answering this question eventually led to what is now called the ``measurement problem'' of quantum mechanics. Namely, it seemed that the properties would become well-defined and thus fully (i.e.\ in the traditional/classical sense) real only upon their measurement, i.e.\ observation. \cor{According to quantum formalism, it was only upon the measurement that the probabilities of various possible results would \emph{collapse} into one truly realized outcome --- a mere potentiality turning into actuality. This idea was often summarized as ``measurement causes collapse of the wavefunction''.}

Yet, what would constitute a measurement? Bohr thought it was sufficient, in this context, to note that to have a (communicable) measurement outcome requires the interaction of the measured system with a macroscopic measurement instrument - where the latter must be described in the language of classical physics \cite{BohrMeasurement}. However, this boundary between a quantum system and the classical measurement apparatus (that possessed this crucial ``ability'' to make system properties well-defined and thus truly real via measurement) - was itself fluid and vague. Indeed, what prevented us from seeing the measurement apparatus as yet another (albeit larger and more complicated) quantum system that adhered too to the principles of quantum mechanics? And if so, then what makes the properties of the apparatus itself (with or without the system it is measuring) well-defined, that is, when \cor{in that case} the reality becomes truly ``real''? Such a chain of thoughts prompted Eugene Wigner to, in a part of his life, even assume that it must be the consciousness of the observer (as purportedly the only entity not describable by quantum laws) which truly completes the measurement process and grants the full reality to physical systems. Needless to say, such a view (if seen, as by Wigner, as a claim that consciousness causes an objective, instead of merely subjective, wavefunction collapse) would open a number of other intractable and presumably ridiculous questions (e.g. could a cat, a mouse, an amoeba, or single neuron cause the collapse?) Overall, to the present day, there has been no consensus about what would be a satisfactory solution to the question when (if ever) the reality becomes well defined (only a plethora of different views called the interpretations of quantum mechanics), so the measurement problem remains.

In any case, Einstein would not subscribe to any of this. The novel roles of the observation and observer in quantum mechanics, to whom was granted the ability to cause the ``wavefunction collapse''\footnote{\cor{I.e.\ to invoke the projection postulate of quantum mechanics.}}, was something abhorred by Einstein. He insisted that physics should describe objective reality independent of any observation, stressing that the moon must be there even when we do not look at it \cite{EinsteinMoon, ClauserReview}). Therefore, he kept attempting to provide arguments that quantum mechanics must be only an approximate description of a maybe complicated but objective and deterministic underlying reality. After each his argument, Bohr would consequently make an effort to refute Einstein's criticism of the theory, with the aim to show that quantum mechanics was not some effective approximation, {\tmpcolor{purple} but} that the quantum wavefunction provides a complete description of reality. These exchanges are remembered as probably the most impressive contest of the two genius minds in the history of science, aka the Einstein-Bohr debates. Einstein would come up with a wisely constructed thought experiment, ostensibly violating the principles of quantum mechanics or demonstrating its logical inconsistency, and then it was up to Bohr to defend the theory by showing where Einstein went wrong. In all rounds of this intellectual game they played, Bohr would eventually come out as the winner (the most remarkable case being one when Bohr used Einstein's general relativity to prove Einstein's argument wrong). Seemingly, Bohr won in all rounds but the last one.

This last and the most influential round of that contest is today known as the Einsten-Podolsky-Rosen (EPR) paradox. It was this argument, and its final experimental resolution that, almost a century later, ended in the last year's Nobel prize in physics -- and therefore it warrants a more detailed discussion.

\subsection{Einstein-Podolsky-Rosen paradox}

In the May of 1935 a paper titled ``Can Quantum-Mechanical Description of Physical Reality Be Considered Complete?'' appeared in the journal Physical Review \cite{EPRpaper}. It was signed by Albert Einstein, Boris Podolsky, and Nathan Rosen. In spite of the question mark in the title, it was clear that the authors were confident that they have finally proved the ``incompleteness'' of the quantum theory. And sure they had some strong arguments.

Effectively, their claim was that the wavefunction -- \cor{the} mathematical object that, according to quantum mechanics, was supposed to represent the complete information about the physical system -- must be actually representing only some ``incomplete'' approximation of the physical reality. Namely, as they argued, in a good and complete physical theory, every ``element of physical reality'' must find its place in the mathematical description of the universe -- but, according to the authors, this was not the case in quantum mechanics. Their point was that likely something is missing from the quantum description, i.e.\ that some properties of physical systems were not taken into account by quantum theory. (Such properties we would, in modern terminology, call ``hidden variables''.) Or, at the very least, that the correspondence between the wavefunction and the physical system is not so one-to-one as claimed by the proponents of the quantum theory (i.e.\ that the theory lacks ``bijective completeness'', as this was later called by some authors \cite{BijectiveComleteness}).

\cor{The novel idea was to consider cases of physical systems that interacted in the past and were then separated. For example, we may consider two particles that constituted a bound two-particle system (like an electron and a proton forming a hydrogen atom) that later decayed, resulting in the two particles leaving in opposite directions. In such cases, it is generally possible to infer some property of one of the particles indirectly, by making the corresponding measurement on the other particle in the pair. For example, if we knew the initial momentum of the overall system (e.g.\ it was zero) and the momentum was conserved, then afterward by measuring the momentum of one particle we immediately find out also the momentum of the other one (i.e.\ it must be of the same intensity but in the opposite direction, so that their vector sum remains zero). This, of course, holds irrespectively of the mutual distance that the separating particles reached at the moment of measurement. In principle, this allows us to find out and precisely predict the momentum of a faraway particle by performing a measurement on a nearby particle.
	
This fact, that we can precisely know the momentum of the distant particle, without ever disturbing it in any way (we have not performed any measurement on it, only on its peer), in the view of EPR authors meant that there must be something \emph{real} about this momentum. Whoever next measures the momentum of that other particle is guaranteed (i.e.\ with 100 percent chance) to obtain precisely the predicted result. EPR paper argues that this is only possible if the distant particle really possesses that well-defined, truly existing momentum value -- in other words, if it is an \emph{element of reality}. Quite generally, EPR paper authors argued that \cite{EPRpaper}: ``If, without in any way disturbing a system, we can predict with certainty (i.e., with probability equal to unity) the value of a physical quantity, then there exists an element of reality corresponding to that quantity''. In a relatively similar way, it turns out to be possible to find out the precise location of the distant particle (by measuring the position of the closer one) which, by the same reasoning, implies that the position of that particle must be a well-defined and real property -- an element of reality. Now, since the two particles can be light years apart at the moment when we perform the measurement on one of them, such measurement cannot immediately affect the distant particle -- therefore, if it has well-defined momentum after we measure the momentum of its peer, then it must have had it all the time. The same with the position. Ergo, contrary to Heisenberg relations and Bohr's interpretation of those, the particle must possess both well-defined momentum and well-defined position (with precise values of both quantities). If quantum mechanics cannot describe both these properties simultaneously, it means that it is incomplete.

In this last part of the reasoning, they tacitly relied on what is commonly called the principle of \emph{locality}, i.e.\ of the idea that no influence can be exerted immediately and directly at a distance (more specifically, that no influence can travel faster than light). Namely, a part and parcel of a fully successful mechanical description of reality was always the idea that matter should affect other matter only ``locally'', that is, only the matter which is in its immediate vicinity (i.e.\ as if by mechanically ``pulling and pushing'') and that only in this way any influence or interaction can propagate through space. It seemed that, at the beginning of the XX century, Einstein cemented this view by showing that the speed of this propagation is always limited by the speed of light (and, via General Relativity, by clarifying that the same is true also for gravitational interaction). 

While this is the outline of EPR's essential argument against the completeness of quantum mechanics, it is worthwhile to consider their discussion in some more detail, especially as a prelude to later developments.

Quite generally, EPR authors have noticed that the type of connection (or, in mathematical terminology, type of \emph{correlation}) that, according to quantum mechanics, often exists between particles (systems) that interacted in the past and were then separated, is very strange from the common-sense classical-physics viewpoint. Actually, as they pointed out, quantum mechanics predicted a whole class of many-particle states with such strange properties, which had no classical-physics analogs.} Today we call such many-particle states ``entangled''. 

\cor{For example, we can consider more closely wavefunction (or, in contemporary terminology, \emph{state vector}, or simply \emph{state}) corresponding to two particles of spin $\frac 12$ which are ``aligned in opposing directions'' in the sense that their total spin equals zero. In other words, while each of the particles separately ``rotates'' in a certain sense -- and thus possesses the axis and direction of this rotation (spin) -- the overall angular momentum (physical measure of rotation) vanishes. This is similar to the case discussed above -- of two particles of known and vanishing total momentum -- only here we consider angular instead of linear momentum, for the reasons of mathematical simplicity. (To practically obtain such a state, the two particles generally had to interact before, to constitute a bound state, or to be simultaneously generated in some physical process.) Mathematically, this two-particle state, in the so-called Dirac notation, has almost an intuitive, pictorial form:	
\be \ket{\Psi} = \frac{1}{\sqrt 2}(\ket{\uparrow}\ket{\downarrow} - \ket{\downarrow}\ket{\uparrow}). \label{EnSpinZ} \ee
Here, $\ket{\uparrow}$ denotes particle spin aligned in the direction of the positive z-axis, i.e.\ ``upwards'' (meaning as if the particle ``rotates'' around the z-axis in anti-clockwise direction as seen from above, where the usual convention is that the z-axis goes in the ``upward'' direction), while $\ket{\downarrow}$ will denote the opposite direction of spinning.} The term $\ket{\uparrow}\ket{\downarrow}$ then corresponds to the physical situation where the first particle has \cor{its} spin oriented upwards, while the second particle's spin has a downward spin orientation. \cor{Hence, the (z components of) their spins cancel each other. The overall state has two similar terms of this type, but the second one is ``opposite'' with respect to the two particles: the first particle is this time spin-down, while the second is spin-up.  To simplify referencing these two particles, we can imagine that the first of these two particles is heading to a researcher named Alice, while the second one is moving towards a researcher called Bob (here we follow the usual naming convention in papers on this subject). Then $\ket{\uparrow}\ket{\downarrow}$ denotes that Alice's particle is spin-up, while Bob's is spin-down and vice-versa for the $\ket{\downarrow}\ket{\uparrow}$ term. The combination of the two terms, together with the relative minus sign, turns out to mathematically ensure that the total spin angular momentum is zero, along any axis (not only along the z-axis). The numerical factor $\frac{1}{\sqrt 2}$ is the so-called overall normalization (in principle, relevant for the evaluation of probabilities) and it is not crucial for understanding the paradox.}

\cor{Taken together, the two terms} have the following simple interpretation: if \alice's particle is spin-up, \bob's will be spin-down, and vice versa. More precisely, the formalism of quantum mechanics tells us a couple more things about the particles in such a state. First, if we measure the spin of \alice's particle along the z-axis, we will get the up result in the 50\% of cases, and the spin down result in the other 50\% of cases. And the same goes for \bob's particle.\footnote{\cor{Mathematically, we calculate the up spin probability of the first particle by computing $\bra{\Psi}\ket{\uparrow}\bra{\uparrow}\otimes I \ket{\Psi}$, where $I$ is the identity operator.}} In addition, the formalism also tells us that if \alice's particle, upon the z-spin measurement, turned out to be spin-up, the measurement of the z-spin of \bob's particle will yield spin down result with certainty, i.e.\ 100\%. It means that, after the measurement of the spin of \alice's particle and getting the up result, the state of \bob's particle is no longer ambiguous (i.e.\ leading \cor{to} up and down outcomes with equal chances), but has become precisely $\ket{\downarrow}$.\footnote{\cor{We can obtain this result by acting with the projector $\ket{\uparrow}\bra{\uparrow}\otimes 1 $ on the state $\Psi$.}} And vice versa: obtaining that \alice's particle has its spin in the down direction, reveals that \bob's particle has its spin upwards -- which in turn means that, as a consequence of the measurement on \alice's particle, the state of \bob's particle was automatically changed, or ``updated'' to the $\ket{\uparrow}$ value.

In this sense, by measuring the z-spin component of the first particle, we have automatically found out also the z-spin of the other particle and, according to the quantum formalism, by the same token, we have also altered (i.e.\ updated) the wavefunction of the other particle spin. And this conclusion holds irrespective of the \cor{distance between} the two particles: even if the other particle is located light years away, by measuring the spin of the first particle we can find out the spin orientation of the second. After all, this is not really surprising: we already knew that the two spins must be opposite, so finding one directly determines the other. In the process, we have updated our knowledge about the spin of the second particle, and as long as its wavefunction corresponds to our knowledge, there is nothing mysterious in the fact that the wavefunction has immediately changed, no matter that the particle is light years away.

None of this so far is particularly revealing. But the key insight of the authors \cor{was} that everything said still holds if we decide to measure the spin along the x-axis, instead of the z-axis. Namely, it turns out that the same state vector $\ket{\Psi}$ can be written, in essentially the same form, this time using ``right'' $\ket{\rightarrow}$ and ``left'' $\ket{\leftarrow}$ spinning states -- with respect to the x-axis:\footnote{\cor{This is not difficult to verify since: $\ket{\rightarrow} = \frac{1}{\sqrt 2}(\ket{\uparrow} + \ket{\downarrow})$ and $\ket{\leftarrow} = \frac{1}{\sqrt 2}(\ket{\uparrow} - \ket{\downarrow})$. Here, the relative minus sign between the two terms in (\ref{EnSpinZ}) and (\ref{EnSpinX}) becomes important.}}

\be \ket{\Psi} = \frac{1}{\sqrt 2}(\ket{\rightarrow}\ket{\leftarrow} - \ket{\leftarrow}\ket{\rightarrow}). \label{EnSpinX} \ee

In particular, this means that if we perform a spin-x measurement on \alice's particle, we will immediately also know the x-spin component of \bob's particle. Furthermore, according to the formalism, it would mean that the state of \bob's particle will be again altered after the measurement performed by {\alice} on her particle: if we obtain that \alice's is right spinning, the state of \bob's will become $\ket{\leftarrow}$, and if we obtain left x-spin of \alice's particle, \bob's will be ``projected'' to $\ket{\rightarrow}$ state.

And now the authors make two important points. The first is that by choosing what measurement to perform on \cor{her} particle, {\alice} can \cor{partially} influence what will be the new state of \bob's particle. Namely, {\alice} can choose whether to measure spin along the z-axis or along the x-axis. In the first case, \bob's particle will end up either in the down state $\ket{\downarrow}$ (if \alice's outcome was up) or in the up state $\ket{\uparrow}$ (if {\alice} obtained \cor{down}). Alternatively, {\alice} can choose to perform measurement of the x-spin \cor{component}, and \bob's particle will end up either in the state $\ket{\rightarrow}$ or $\ket{\leftarrow}$. Note that {\alice} cannot choose whether \bob's particle will end in $\ket{\rightarrow}$ or in $\ket{\leftarrow}$ state -- this is completely random, with probabilities 50 \% for both outcomes. But \cor{she} can choose whether \bob's particle will end up in the \cor{set} $\{\ket{\uparrow}$, $\ket{\downarrow} \}$ (if \cor{she} decides to measure spin along the z-axis) or in the \cor{set} $\{\ket{\rightarrow}, \ket{\leftarrow} \}$ (if \cor{she} opts for the x direction). And \cor{she} can decide in which of these two \cor{mutually exclusive sets} \bob's particle will turn up after \cor{her} measurement, even if \bob's particle is light years away \cor{at that moment}!

This {\tmpcolor{purple} seemed clearly} at odds with Einstein's theory of relativity, according to which no influence {\tmpcolor{purple}could} travel faster than light. {\tmpcolor{purple} As it is widely known, light speed appears as the natural speed limit in our universe, since the theory of relativity has shown that a bunch of problems and paradoxes are related to exceeding this limit. On top of the list are problems with causality. Namely, thanks to Einstein's discoveries, we know that moving faster than light in one system of reference amounts to traveling into the past in another system. And while wishful thinking on behalf of the science fiction industry has made worldwide audiences pretty comfortable with the time travel idea (as long as one takes care not to kill own grandfather), in reality, it is the logical inconsistencies that plague (and thus prevent) this hypothetical possibility. The mere possibility to (on demand) send information into own past is a logical contradiction: one can decide to send oneself in the past (e.g.\ yesterday) any piece of information that he has not already received yesterday -- proving that this is either impossible or not deserving to be truly called "sending information into past". We emphasize that the problem has nothing to do with free will -- as long as a deterministic robot has the means to send arbitrary information to its own past, a simple algorithm of this sort demonstrates the implausibility of the idea.\footnote{\cor{It should be noted, however, that the existence of so-called ``closed time loops'', as well as tachyons, is not explicitly forbidden by contemporary physical theories, but to avoid logical contradiction the related hypotheses must somehow preclude either sending information into the past altogether, or at least the ability to do it at will.}} For these reasons, it is -- and was at the time of EPR writing -- generally taken that no influences can travel faster than light.}
	
Thus, the authors {\tmpcolor{purple}of the EPR paper drew} a seemingly natural conclusion: since no real influence can travel faster than light, it must be that nothing \emph{really} happens with \bob's particle upon the \cor{distant} spin measurement performed \cor{by Alice}. Consequently, it means that some of the states from the $\{\ket{\uparrow}, \ket{\downarrow} \}$ set must correspond to the same unchanged reality of \bob's particle as one of the states from the $\{\ket{\rightarrow}, \ket{\leftarrow} \}$ set. That is, the state vector (wavefunction) has changed, but the reality has not -- hence this matching between the wavefunction and the true reality of \bob's particle is not so one-to-one. In other words, the quantum wavefunction does not provide such a completely adequate description of reality as Bohr liked to claim. Either that, or there was some ``spooky action at distance'', as Einstein liked to call this immediate influence that the measurement on the first particle allegedly has on the second one. {\tmpcolor{purple} For Einstein, the latter was hardly a serious option.}

At this point, it is important to understand why authors could not make a stronger claim: that, since the quantum theory predicts such immediate action at a distance, it is in strict violation of special relativity (and that, thus, quantum mechanics is simply wrong). Namely, {\tmpcolor{purple} one needs to be able to send information faster than light at will, to violate the predictions special theory of relativity: according to relativity, by consecutively sending information faster than light and then returning it back again faster than light from another system of reference, one can actually send it in own past -- which, we explained, is not simply weird but is logically forbidden. But, curiously, in spite of allowing the existence of this ``spooky action at distance'' that influences wavefunction}, quantum mechanics predicts that sending information via such a channel is not possible. This has to do with the inherent randomness of quantum mechanics, due to which \cor{Alice cannot influence in which particular} state vector will \bob's particle end (i.e.\ \cor{she} can choose only in which of the two sets it will end). And the quantum calculations show that {\bob}, by measuring the spin of \cor{his particle cannot infer whether Alice has measured her spin along the z-axis or along the x-axis. (All Bob can do is to measure his particle spin along some axis, but irrespective of Alice's choice, his spin measurement will always yield 50-50 probability along any axis.)} Therefore, \alice's decision on whether to measure x-spin or z-spin \cor{of her} particle cannot be later inferred just by performing measurements on \bob's particle, in spite of the fact that its wavefunction description will be different in the two cases. 

{\tmpcolor{purple}
Had the quantum mechanics predicted that entangled pairs can be used for sending faster-than-light signals (obviously then, it would allow sending information ``on demand''), this would have been in contradiction with the implications of special relativity. In principle, one could then perform an actual experiment and check whether the superluminal signaling in this way is really achieved -- falsifying one or the other of the two theories. But this was an awkward situation - according to the wavefunction description, some influence was spreading faster than light, yet still it was impossible to use such influence to send meaningful information. As if the two particles were somehow communicating faster than light, only in some specific conspiratorial way that we cannot use for sending information ourselves. Something like that was unconceivable in classical physics, but was somehow now strangely allowed in quantum physics, thanks to its indeterminism. This might have contributed to Einstein calling the effect ``\emph{spooky} action at distance'', instead of simply ``action at distance''. Consequently, the EPR paradox could not demonstrate the inconsistency of quantum mechanics nor that it has any real (i.e.\ measurable) conflict with relativity, and for this reason, the EPR paper remained only a (strong) philosophical argument against the purported completeness of the quantum theory. On the other hand, for Einstein, who was a realist and did not believe that nature is truly indeterministic, whether \emph{we} could use this supposed influence to send information or not certainly did not seem essential: in his view, no real influence could travel faster than light and thus the fact that, according to quantum mechanics, wavefunction was nevertheless changing at distance was simply a proof that quantum description was faulty. }

The second point the EPR authors made \cor{was the one related to uncertainty relations, that we have already briefly discussed}. Namely, Heisenberg relations also hold for x and z components of spin.\footnote{\cor{The $S_x$ and $S_z$ operators, corresponding to measurements of these two values, do not commute.}} This means that, according to Bohr's understanding of the uncertainty relations, when spin along the x-axis is precisely known, the spin along the z-axis cannot be well defined, and vice versa. \cor{But,}  according to Einstein, Podolsky, and Rosen, this two-particle setup demonstrates that Bohr's understanding must be wrong, that is, that values of spin along both axes must be well defined and existing properties \cor{at all times}, while it is only due to the imperfection of the quantum theory (i.e.\ due to its incompleteness) \cor{that} these properties do not appear in the mathematical formalism.

Their argument goes as follows. Let us assume that {\alice} first measures spin \cor{of her particle} along the z-axis, and obtains the up result. Now we know (and quantum formalism tells us) that if \cor{Bob measures the z-spin of his particle, he is} bound to get the spin-down result, with certainty. And, according to the authors, since the z-spin of \bob's particle is now well defined, and has the certain, known-in-advance value, it must be that this spin-z property is in some sense real, that is, there must be really ``something about'' \cor{this} particle that would ``produce'' that guaranteed outcome upon future z-spin measurement. Furthermore, since these particles can be arbitrarily distant from each other (precluding any \cor{real} influence of \alice's measurement on \bob's particle), that property must have been there even before \alice's measurement or, more precisely, irrespective of any measurement. Thus, this property of \cor{Bob's particle} that determines \cor{its} z-spin outcome must be there even if {\alice} decides not to carry out any measurement. \cor{Thus, authors call this real property of Bob's particle ``an element of reality'' corresponding to its z-spin value.}

By the same reasoning, we can conclude that there must be some feature of \bob's particle -- i.e.\ \cor{an} element of reality -- that determines the outcome of the x-spin measurement, since its x-spin value we can also, in principle, infer by measuring the spin of \alice's particle along the x-axis. In this reasoning, it is crucial that we are able to reveal the spin of \bob's particle without influencing, i.e.\ disturbing this particle, as it is here guaranteed by the spatial distance between the particles \cor{ -- of course, assuming the validity of the ``locality'' premise (which was never doubted in the paper). Namely, if any influence of Alice's measurement on Bob's particle was possible, then there would remain a possibility that the spin value of Bob's particle became well-defined only after Alice's measurement and due to it. This is why the ``without in any way disturbing a system'' requirement had to be present in the above EPR definition of an ``element of reality''.} 

In this way, the authors have concluded that there must exist elements of reality that determine the spin projection both along the x-axis and along the z-axis, and since these do not appear in quantum formalism, this formalism must be incomplete. While this is the entire formal conclusion of the paper, essentially the authors expected that they have consequently proved that uncertainty relations are therefore a consequence of this incompleteness, i.e.\ of our technical inability to precisely infer these elements of reality by measurement and account for them in the theory. In other words, reality is rock-solid with always well-defined properties, while the problem is in us and our present theory. Such elements of reality that do not show up in quantum mechanics, due to its alleged incompleteness, soon become known (in general) as ``hidden variables''. \cor{All this was} contrary to Bohr's stance that quantum mechanics provided the full picture (with no room for any ``hidden variables''), while the x-spin value, after the spin z measurement, simply was not a well-defined property of nature, per se (following the quantum formalism, this property existed then in a \cor{so-called} superposition of different possibilities, and thus did not have a well defined unique value).

The ball was now in Bohr's court. It took him a few months to reply, and he did it by publishing a paper with precisely the same title as the title of the EPR paper, in the same journal \cite{EPRreply}. {\tmpcolor{purple}As we will discuss in some more detail later, the reply emphasized his ``non-realist'' views (in a certain sense: if something does not exist well defined per se, what would it mean that it was changed at a distance) and that we must embrace a new view of reality, which is more holistic (instead of reductionistic) and does not allow us to see the two particles as truly separate entities.} Until this day there are ongoing discussions of historians of science and philosophers to what extent his defense of the completeness of the theory was to the point and successful, e.g.\ \cite{EPRStanford, EPRreplyDisc1, EPRreplyDisc2, EPRreplyDisc3, EPRreplyDisc4, EPRreplyDisc5}. Many of the analyses are of the opinion that the reply was unclear and confusing, and even Bohr himself, some 15 years later, has regretted his choice of wording on some essential points \cite{EPRStanford}. However, what was certain is that he doubled down on the completeness of quantum mechanics, even if it entailed accepting some sort of nonlocality, i.e.\ of the ``spooky action at distance''.

In any case, both the EPR's initial argument and Bohr's reply were of philosophical nature, so it was rather a matter of taste who will find which side more appealing and convincing. And the \cor{conundrum} seemed likely to remain undecided, maybe forever.

\section{John Bell}

\subsection{Hidden Variables}

For the second act of the drama, we fast forward some thirty years into the future.

The success of the new physics was huge, and almost nobody really cared about Einsten's philosophical objections. In the meanwhile the formalism was polished (mostly thanks to John von Neumann) and in practice it worked flawlessly, showing no traces of any incompleteness. The measurement problem still existed, but was not showing up in any real-life applications or computations. Besides, the post-WWII era science was more of a pragmatic type, and physicists were increasingly leaning towards the infamous ``shut up and calculate'' view of quantum mechanics.

One of \cor{the few} to whom the question of the true meaning of quantum mechanics was of greater importance than its applications, \cor{was} Irish physicist John Bell. Working on accelerator design at CERN in the early sixties, Bell was still managing to find time to investigate the deepest interpretational issues of quantum physics.

Personally, Bell was entirely sharing the late Einstein's dissatisfaction with quantum theory. Even the quantum terminology alone was, in his view, improper for a theory of physics -- as we can see from his comment about the books on the subject \cite{BellSpeakable}:
\begin{quote}
'For the good books known to me are not much concerned with physical precision. This is clear already from their vocabulary. Here are some words which, however legitimate and necessary in application, have no place in a formulation with any pretension to physical precision: system, apparatus, environment, microscopic, macroscopic, reversible, irreversible, observable, information, measurement. ... On this list of bad words from good books, the worst of all is ``measurement''.'
\end{quote}
So, the roles of the measurement and the observer in quantum mechanics, even more than the lack of determinism, were deeply unsettling to him, similarly as they were to Einstein. In his view, the prevailing Copenhagen interpretation of quantum physics (i.e.\ one following the views of Bohr and Heisenberg) had too many ``subjective'' elements, while he longed for a description of objective reality that would be devoid of vagueness, indeterminism, and role of observers \cite{BellSpeakable}.

In particular, he was very enthusiastic about the so-called de Broglie-Bohm pilot-wave interpretation of quantum mechanics. \cor{According to the pilot-wave theory, particles had well-defined positions at all times and obeyed causal (deterministic) laws. These supposedly real and well-defined positions were an example of what EPR authors considered as elements of reality that were not taken into account by the standard quantum mechanics. In quantum mechanics, the full description of systems was given by the wavefunction, and wavefunction only, so if these well-defined coordinates really existed, they were ``hidden'' in the quantum formalism. Therefore, the pilot-wave theory was an example of a model with \emph{hidden variables} that was nevertheless experimentally indistinguishable from the standard, Bohr's quantum mechanics.}

{\tmpcolor{purple} However, while realistic and deterministic, this model had a terribly unwelcome feature of being manifestly nonlocal -- meaning that influences could spread immediately over distance, as if acting at infinite speeds. In particular, the velocity of a particle in this theory would depend on the configuration of arbitrarily distant particles at the present moment, i.e.\ in principle it could depend on what was currently happening in other galaxies.} Due to special relativity, this meant that the influences could also travel into the past, as seen from some reference systems. In truth, the pilot-wave model predicted that information could not be sent faster than by light speed, so it did not experimentally conflict with relativity (otherwise it could be easily falsified), but the nonlocality feature also led to many other, either technical or conceptual complications that plagued this approach. Nevertheless, for Bell, this was an encouraging example, a proof of concept that it might be possible to find a truly satisfactory realistic underlying theory that would replace the vague and subjectivistic quantum formalism. In particular, it would have been exactly the type of theory Einstein was arguing for in the EPR paper, if only it was not for this nonlocality. The essential question was could a similar realistic, but this time also local theory exist, and how to find it?

Bell was strongly inspired by the EPR paradox. In the meanwhile, the experimentalists managed (in experiments with photons) to check that the entangled pairs of particles indeed behaved as quantum mechanics predicted: the two spins in the state given by Eq.\ (\ref{EnSpinZ}) would indeed always be opposite to one another (along arbitrary direction). Nevertheless, this confirmation did not take anything away from Einstein, Podolsky, and Rosen's arguments of incompleteness (after all, they exactly presumed that spins would always be opposite, in deriving their conclusions). Just like Einstein, Bell was confident that the fact \cor{that} two spins always turn out to be opposite, across arbitrary distances, must point to the existence of some well-defined underlying ``elements of reality'' that were driving such experimental outcomes. Basically, he understood that these spins must in some sense simply be real, predefined, and opposite to each other, and that the experiment's role was just to confirm and demonstrate such a state of affairs, and not to \emph{make} any of the spins well defined at the moment of measurement (as quantum formalism suggested). He likened the case of opposing spins in the EPR pair to the socks of his colleague Reinhold Bertlman, who had a habit of wearing socks of different colors. In his paper ``Bertlmann's socks and the nature of reality'' \cite{Bertlman}, he noted that when we observe one sock of Prof. Bertlman we immediately know that the other sock will be of a different color, even before we spot it -- and there is nothing puzzling here \cor{(apart from Bertlmann's dressing habits),} irrespective of the hypothetical distance between the socks. The colors of both socks were real and well defined all time, and in no way the moment of our observation was essential nor influenced any of the socks. He was positive that the matter of spins should be no more mysterious than that of his colleague's socks, once we discover the underlying true theory of quanta.

Yet, there was this question of how to prove that the spins were indeed like socks and that measurement outcomes were merely revealing the preexisting well-defined reality? Thus far, the EPR argument had remained purely a philosophical one and there had been no consensus about its implications, let alone a consensus that a more complete theory than quantum mechanics was really needed. Moving forward from there required an argument in a form of a prediction that could be experimentally verified and which would, once and for all, settle down the dispute either in Einstein's favor or in favor of Bohr.

However, this seemed impossible. Namely, the trouble was that nobody yet knew the hypothetical underlying more-complete theory that Einstein was arguing for. And if we do not know the theory, how could we possibly test and compare its predictions? All anyone knew was that such a theory had to be realistic and local, but apart from that, it could be arbitrarily complex. Einstein was not claiming that he knew it -- he was only confident that it must exist, but the theory might in principle be even so complicated that human beings could be incapable of deducing it. \cor{Besides, if existed, why would such a theory experimentally differ at all from quantum mechanics -- maybe all experimental predictions would be the same, as was the case with the pilot-wave theory?} 

But, Bell's intuition was different. He felt that any such locally-realistic theory must be observably different from quantum mechanics, and in 1964 he \cor{managed to} prove it \cor{\cite{Bell64}}. The proof is known as Bell's theorem, and it (together with the associated experimental tests) was even praised as ``the most profound discovery of science'' \cite{StappOnBell}. Note that the claim was not about the most profound discovery of \emph{physics}, but of all \emph{science}, in the history of mankind!

\subsection{Bell's theorem}

The starting point was the setup with two entangled particles depicted in the EPR paper. Bell felt that the type of connection between the particles predicted by quantum theory was in some sense ``stronger'' than it was possible to explain by any ``pre-existing'' reality (such as that of well-defined pre-existing colors of Bertleman's socks). It did not take long to realize that, to prove this, it was not sufficient to consider only the type of measurements that were in the spotlight of the EPR paper -- that is, those where the outcomes of the measurement on the second particle were \cor{certain}. Thus, Bell's first insight was that he should in much more detail consider non-zero-or-one probabilities related to measurements on both particles, along different directions.

Once he realized this, he went on to derive a mathematical relation concerning a combination of probabilities to get particular spin outcomes on the two distant particles, but when we are measuring the spin along axes that are neither parallel nor perpendicular, but forming some intermediate angles. For example, he was considering probabilities of the following type: if \cor{Alice measures her particle spin along z-axis and obtains up}, what is the probability \cor{for Bob} to obtain the spin of \cor{his} particle to be in the direction inclined by 45 degrees with respect to the vertical? He managed to mathematically prove that in \emph{any} locally-realistic universe, a particular combination of such probabilities \emph{must} satisfy a certain inequality, now known as Bell's inequality. The main achievement here was in this word ``any''! This meant that truly \emph{any} universe that would respect Einstein's (and Bell's) expectations -- i.e.\ where physical systems have well-defined properties irrespective of observation and where influences cannot spread infinitely fast -- was bound to satisfy \cor{this inequality}. This included arbitrarily complicated universes, where particles could in principle mutually communicate and even conspire against experimentalists in arbitrary manners, or could behave differently on Earth than on Jupiter (whether we as humans are able to infer or comprehend these complex rules or not, was of no relevance here). 

\cor{It is important to underline that this inequality had, per se, absolutely nothing to do with quantum mechanics. Again, this was inequality that necessarily constrained any theory in which measurement outcomes were driven by observer-independent well-defined properties of matter (i.e.\ adhering to realism) and in which there was a limit for the speed at which influences can travel (i.e.\ respecting locality). In other words, this was a limitation for all universes that meet Einstein's expectations, not for the quantum mechanical universe of Niels Bohr. 
	
This is clear from the fact that the inequality was essentially derived from the following two seemingly naive and self-evident, presumptions. The first was the premise of locality: whatever Alice decides to do in her lab, it cannot influence the outcomes (or probabilities of outcomes) of Bob's simultaneous measurements in his arbitrarily distant lab. The other was the premise of realism, which was, arguably, tacitly always present in derivations. In the original Bell's formulation, this premise was taken into account by assuming that the outcomes of spin measurements deterministically followed from some unknown (and non-existing in quantum formalism) yet well-defined properties of particles. In other words, that the outcomes were determined by some hidden variables. Mathematically, they were explicitly represented in \cite{Bell64} by variable(s) denoted as $\lambda$. In some of the following reformulations of Bell's theorem \cite{BellStanford}, the requirement of determinism was removed, but the premise of realism nevertheless remained in some form through the role of the variable(s) $\lambda$ still representing the hidden variables and through the way these hidden variables were used to derive probabilities of spin measurement outcomes.\footnote{In \cite{BruknerReality} it is argued that the very way the probabilities are expressed as functions of $\lambda$ encodes the presumption about a classical-like, non-contextual reality. Another way to see this is that, arguably, certain presumptions about underlying reality are necessary even to define the locality postulate in the sense in which it is different from ``no-signaling'' (while the former is claimed to be violated, the latter is not).}  

To fully understand how it was possible to prove such a general inequality can be accomplished only by considering the proof itself. However, since the exposition of the proof requires some equations and some (though fairly basic) mathematical knowledge, we discuss it in a separate Appendix \ref{app:inequality}, below. 
}

The entire relevance of this inequality for the context of the EPR discussion of quantum physics was that Bell noticed that quantum theory actually predicted violations of this inequality! \cor{(In the Appendix \ref{app:violation}, we give a concrete mathematical example of this violation.)} Hence, this was precisely what could settle Einstein-Bohr debate once and for all. This inequality was violated by physics advocated by Bohr, while it had to be satisfied by any type of physics advocated by Einstein. Whether the real \cor{physical} systems satisfied this inequality or not could finally discriminate between Bohr's and Einstein's views, and thus this inequality had the power to move the debate from pure philosophy into physics.

The very fact that such inequality existed was proof of Bell's theorem, which essentially stated that no locally-realistic theory can ever reproduce exactly all quantum mechanical predictions. In other words, this was also proof that it was not possible to improve the pilot-wave theory in a way to make it local but still experimentally indistinguishable from the standard \cor{quantum} theory. It was now certain that any realistic model that closely reproduces experimental predictions of quantum formalism is also bound to have non-local properties. And it is important to stress that this conclusion holds irrespective of whether we are looking for a strictly deterministic theory or we are even ready to accept inherent randomness (this, however, had become clear only after some generalizations of the original Bell's derivation).

Crucially, the quantities constrained by Bell's inequality were something that in principle could be measured. Therefore, it finally became possible to experimentally measure values of these quantities -- for some of the special cases in which quantum physics predicts violation of the inequality -- and simply check if the constraint is satisfied or not. If the inequality would turn out to be preserved in such an experiment, that would be in contradiction with quantum predictions and would present definite proof that quantum mechanics requires modifications. That would mean the entire quantum theory was only some sort of approximation, with just partial validity. Bell, himself thought that such an outcome was, regrettably, unlikely (discouraged by the enormous practical success of quantum mechanics), but still had a hope of ``an unexpected result, which would shake the world!'' \cite{ClauserReview} and falsify the completeness of quantum ideas.

However, if the measurement would reveal a violation of Bell's inequality, it would have groundbreaking philosophical implications for our understanding of the universe. Once and for all we would have to abandon any hope that a ``rational'' realistic explanation of phenomena in nature, of the type sought after by Einstein, could be found.

Bell, for his part, could now just sit and wait until someone managed to perform the real experiment.

\section{John F. Clauser, Alain Aspect, and Anton Zeilinger}

\subsection{Experimental verdict}

The history of physics had to wait almost another decade for the final act of this philosophical drama. Years were passing by after Bell's groundbreaking proof, yet hardly anyone was noticing Bell's discovery. Bell had to wait a couple of years even for the first citation of his work, collecting only a handful of citations in the first decade or more. It was not only that his paper was published in a less-known journal ``Physics, Physique, Fizika'' \cite{Bell64} but, as the young physicist John F. Clauser working as a postdoctoral researcher at UC Berkeley at the time would later recall, this was an unpopular topic at all. Young Clauser, together with his colleague, graduate student Stuart Freedman, will be the first ever to perform the experimental test of Bell's inequality. Even after being warned that this research might ruin their careers \cite{ClauserInterview, ClauserReview}, the two young scientists pursued further their idea to finally obtain a definite experimental answer about who was right in the great Einstein-Bohr debate.

From 1964 to 1972 our understanding of physics stayed in an unusual limbo state: we knew it was possible to, once and for all, find out whether our universe is ``normal'' in the sense of Einstein (local and real), or truly as ``weird'' as Bohr suggested -- but still nobody knew the answer. And finally, in 1972, Clauser and Freedman managed to perform the first ever ``Bell's test''. \cor{Unlike the original Bell's paper \cite{Bell64}, which discussed matter particles of spin equal to $\frac 12$ (such as electrons or protons), for practical reasons this experiment dealt with photons, measuring polarizations of photon pairs emitted in an atomic cascade \cite{ClauserExperiment}. While the spins in Bell's analysis were opposite, here the polarizations of the two photons in each pair were the same: every time we would let the two photons arrive at polarizing filters aligned in the same direction, either both photons would pass the filters, or neither would. The task was to measure so-called coincidences: how often would it happen that both photons managed to pass the corresponding polarizing filter and to investigate this probability as a function of the angle between the polarizers. Instead of the probability of the spin measurement outcomes, it was these probabilities of photon coincidences that had to satisfy Bell's inequality. However, these were merely technical differences -- the essence remained absolutely the same. 

And,} at last, we heard Nature's ruling -- \cor{no} doubt, it was in favor of Bohr! The obtained experimental results perfectly followed quantum mechanical predictions, demonstrating Bell's violation at 6.3 sigma (where everything over 5 sigmas is considered in physics as statistically sure proof). Clauser, himself sympathetic to Einstein's side in the debate, will later say ``I was very sad to see that my own experiment had proven Einstein wrong'' \cite{SadForEinstein}. But so it was.

No doubt that Bell was even more sad than Clauser. As most dramas have some tragic character, in this one that role was, so to say, taken by Bell. It was Bell's strong inclination towards the realist stance, that led him to investigate possibilities for local hidden variable theories in the first place, finally leading him to his theorem. No one else has so openly and bluntly expressed his convictions as did Bell, and seemingly no one's convictions were as strong as Bell's. He was absolutely sure that Einstein must be right, yet he had to concede that he was not \cite{BellQ1}:
\begin{quote}
For me, it is so reasonable to assume that the photons in those experiments carry with them programs, which have been correlated in advance, telling them how to behave. This is so rational that I think that when Einstein saw that, and the others refused to see it, he was the rational man. The other people, although history has justified them, were burying their heads in the sand. I feel that Einstein’s intellectual superiority over Bohr, in this instance, was enormous; a vast gulf between the man who saw clearly what was needed, and the obscurantist. So for me, it is a pity that Einstein’s idea doesn’t work. The reasonable thing just doesn’t work.
\end{quote}
On another occasion, he succinctly summarized his sentiment as: ``Bohr was inconsistent, unclear, willfully obscure and right. Einstein was consistent, clear, down-to-earth and wrong'' \cite{BellQ2}.

In spite that his theorem and the subsequent violation of the inequality were terribly disappointing to Bell, this episode was a perfect testimony to the strength of the scientific method. While every scientist must and should have their own opinions, a truly good scientist will not allow their own wishful thinking to stand in the way of reaching the truth. In spite of so fervently advocating Einstain's positions, it was essentially Bell himself who led us to the definite falsification of these views. While Clauser was an accomplice, it was Bell who killed the worldview he cherished as the true one. And he did not hesitate to acknowledge that the outcome of his work and of later experiments was not the one he was hoping for. By doing it, Bell established his greatness as a scientist, in yet another way.

\subsection{Closing Loopholes}

Yet, some hope for Einstain's viewpoint still remained after the 1972 experiment. Namely, in Clauser and Freedman's 1972 experiment, the photons were impinging on the polarizers that were a mere couple of meters away, in the same lab. The experiment was performed in cycles of 100 seconds of accumulating the detectors' data, during which the polarizers were kept oriented along fixed angles \cite{ClauserExperiment}. But the essence of Bell's original idea was that the two polarizers belonging to Alice and Bob need to be sufficiently distant, so that Alice's decision about how to orient her polarizer could not \cor{have chance to} influence the photon arriving on Bob's side. In Clauser's experiment, the hypothetical influence of the decision on how to orient the polarizer on one side of the lab had plenty of time to reach the other side, with no need for super-luminal speeds and conflict with locality principles of special relativity.

Reading the original Bell's paper, there is an impression that Bell himself has not even hoped that the outcome of an experiment such as Clauser's could be any different than to support quantum mechanical predictions. On the other hand, his hopes were that quantum mechanical predictions about entangled pairs of particles may be limited to the cases where the sub-luminal influences \cor{can} arrive from one side to the other in due time, but may fail in cases when the distances are greater (or relevant times shorter) \cite{Bell64}:
\begin{quote}
''Of course, the situation is different if the quantum mechanical predictions are of limited validity. Conceivably they might apply only to experiments in which the settings of the instruments are made sufficiently in advance to allow them to reach some mutual rapport by exchanging signals with velocity less than the speed of light. In this connection, experiments of the type proposed by Bohm and Aharonov, in which the settings are changed during the flight of the particles, are crucial.'' \end{quote}
In the last sentence, we see that he urged for experiments that would remove this possibility of having any ordinary, slower-than-light signals influencing the results. 

However, it must be noted that it is difficult to imagine in what way the two polarizers/detectors from different sides could ``communicate'' in order to fool the experimenter. Clauser, personally, saw this as a form of ``paranoia'' for which ``it is first necessary to believe that a pair of detectors and analyzers that are several meters apart are somehow conspiring with each other, so as to defeat the experimenter'' \cite{ClauserReview}. Nevertheless, this remained a logical possibility, so it is generally accepted that Clauser and Freedman's original experimental realization with static polarizers suffered from this ``locality loophole'', as this ``paranoid possibility'' became known.

The first experiment to almost entirely close this locality loophole and essentially meet Bell's original criterion was performed in 1982. It was done by \cor{our second Nobel laureate}, Alain Aspect, and his collaborators working at École supérieure d'optique in Orsay \cite{AspectExperiment}. Among other experimental improvements, Aspect has found a way how to effectively change the orientation of the polarizers on the fly, i.e.\ while the photons were on their way towards detectors. The polarizers (with the corresponding detectors behind) were situated on the opposing ends of the lab, at a distance of 6 meters from the photon source (i.e.\ 12 meters apart). That corresponded to the flight time of 20 nanoseconds from the creation of photons to their detection. Instead of physically rotating the polarizers -- which was technically impossible to accomplish in such a short time -- he replaced the polarizer on each side with a setup consisting of a so-called optical switch and two polarizers aligned in different directions. The switches allowed the incoming photons to be directed towards different polarizers, and this change of direction was occurring regularly, approximately every 10 nanoseconds. In this way, the photons were arriving at the polarizers whose angle varied, and this variation of the angle was happening during the flight of photons. There was no way for a signal, traveling from one side to the other at a speed not greater than that of light, to arrive in time and ``inform'' the physical system there about the current state of the optical switch. The results were unwavering, confirming Bohr's expectations at 5 standard deviations. Bell's last hopes that our universe might still be locally realistic were dispersing.

However, this was not yet precisely what Bell envisaged. Ideally, the experimenter on one side should be capable to orient her polarizer at will, or randomly, so that there would be absolutely no way for the system on the other side to predict what angle was chosen. This was not what Aspect's setup allowed, where variations of polarization were periodic, and thus, strictly speaking, easily predictable (if the equipment on the two sides could communicate, wanted to ``conspire'' against the experimenter -- as Clauser would say, and \cor{was} ``intelligent'' enough to get the pattern of variations). Whether this was an overly paranoid requirement or not, yet it was a logically necessary one if we wanted to be absolutely sure that no room for any locally realistic model remained. And the first one to definitely close this locality (or communication) loophole -- by using superfast random number generators -- was \cor{the third laureate} Anton Zeilinger with his team from the University of Innsbruck, back in 1998 \cite{ZeilingerBell}. His experiment demonstrated a violation of Bell's inequality to a staggering 30 standard deviations.

But, after careful analysis, physicists have realized that, strictly speaking, the locality loophole was not the only potential loophole. Additionally, they identified also the detection loophole (related to the low efficiency of used detectors), the coincidence loophole (related to the imperfect methods used to judge which two detected photons belonged to the same pair), and the memory loophole (a pretty ``paranoid'' possibility that hidden variables could exploit the memory of past measurements to conspire against experimenters). However, with a combined effort of many experimental groups worldwide, finally, in 2015 we had the first ``loophole-free'' confirmations of Bell's inequality violation (one of three such experiments done in 2015 was performed again by Zeilinger's group \cite{LoopholeFree}). 

{\tmpcolor{purple}It is also worth mentioning that some of these additional loopholes do not even appear in generalizations of Bell's theorem to entangled states of three or more particles. Generalizations of this sort were discovered by Daniel Greenberger, Michael Horne, and Anton Zeilinger in 1989 (aka GHZ theorem \cite{GHZtheory}), and the corresponding experimental tests, performed also by Zeilinger \cite{GHZexperiment}, again falsified local realism. In these cases, the incompatibility of local realism with our universe could be verified on the level of a single measurement, without the need to gather sufficient statistics and consider any probabilities, or inequalities at all. And, after all these experiments there was no more any speck of doubt: our world could \emph{never} be explained by any locally realistic theory, and it was Bohr who won the famous debates.}

Overall, hardly to anyone's surprise, the loopholes failed to save local realism. Years later, it is time to finally get used to this fact and take the conclusions seriously.

\subsection{Superdeterminism}

It should be noted that, in a tendency to give statements of utmost mathematical precision, physicists often mention yet another hypothetical loophole. It is known by the name ``superdeterminism''. If the locality loophole was already paranoid in Clauser's words, this one must be labeled as ``insanely paranoid''. Namely, what if the conspiracy of nature (to fool physicists and leave a false impression of Bell's inequality violation) is so grand, that neither random nor human choices for polarizing angles are sufficiently independent? In principle, the universe could still be real and local, moreover deterministic, only if it is precisely fine-tuned at the moment of the Big Bang so that each time we try to test Bell's inequality our attempts are actually futile, since everything is staged. Everything is so precisely adjusted in advance (and the deterministic assumption in this picture makes such an idea logically consistent) that each time we attempt to measure distant particles in a Bell's pair, Bob's particle and/or equipment ``knows'' what angle Alice is going to choose not because it was somehow signaled/influenced faster than light, but simply because everything is evolving according to a script written long ago, by a screenwriter whose core motive seemingly was to trick physicists into a false belief.

John Clauser, together with Abner Shimony and Michael Horne (scientists who, together with Richard Holt, were responsible for the most popular CHSH variation of Bell's inequality) in their 1976 paper \cite{ClauserSuperdeterminism} illustrated this type of conspiracy ideas by noting a possibility that manufacturers of the measurement equipment might have tweaked the instruments as to produce a prepared set of false experimental outcomes. However, for this to work, conspirators also had to instruct the secretaries of the experimenters on both sides to ``quietly whisper in their ears'' appropriate choices of polarizer angles for each run -- everything carefully orchestrated in order to demonstrate the ostensible violation of Bell's inequality in a universe which is otherwise locally real. Actually, in the meanwhile, it turned out that even this scenario would not have worked, since all the details of the conspiracy had to have been fixed much earlier, probably already at the dawn of time: a Bell's test performed in 2018 has confirmed the violation by using, instead of random generators, the light emitted 7.8 billion years ago from distant quasars, to decide how to orient polarizers (so, according to superdeterminism, these quasars would also have to be a part of the conspiracy, taking care of what light to emit, so that some scientists on Earth 7.8 billion years later will be properly fooled) \cite{QuasarsTest}. Furthermore, ``as a safeguard against potential systematic or conspiratorial effects'', one of the loophole-free Bell tests in 2015 \cite{MontyPythonTest} used a number of popular culture sources to modulate the decision of the angles to be used in the experiment. In particular, they used binary bits of digital versions of movies such as the Back to the Future franchise (all three parts, of which the third was used in reverse order), Star Trek episodes, Doctor Who, and the ``Monty Python and the Holy Grail''. For good measure, they also performed a binary XOR operation of everything also with binary digits of the number pi. As explained in \cite{BellStanford}: 
\begin{quote}
``A hypothetical cause that achieved the observed statistics via correlation between states of the photons studied and the choice of measurements would have had to precisely orchestrate the creative processes leading up to the digitized versions of these cultural artifacts in such a way that, when processed in conjunction with the outputs of the random number generators, produced just the right sequence of experimental choices.''
\end{quote}

By one of the worse misnomers in physics, the negation of the ``superdeterminism'' hypothesis is sometimes called the ``free will'' requirement, potentially misleading audiences that the superdeterminism conspiracy idea has anything to do with the ancient and open philosophical problem of the existence of free will. While the existence of free will (or of true inherent randomness in nature) would indeed logically ensure that superdeterminism is not possible, the lack of free will by no means implies superdeterminism. As should be clear from the above discussion, apart from complete determinism (and thus the absence of free will), the latter seemingly also requires some sort of intentional ultra-fine tuning specifically aimed at tricking physicists interested in Bell inequalities. (Here we discuss the superdeterminism in its basic form -- e.g. there are also variations that presuppose retrocausality, i.e.\ affecting past from the future, but this has some resemblance to the idea of putting out fire with gasoline: a rather problematic and hard-to-motivate move to avoid the milder initial problem of Bell's nonlocality.)

As an idea of conspiracy without bounds, the superdeterminism loophole is, obviously, in principle experimentally unfalsifiable. However, if superdeterminism is true, the inability to draw correct conclusions from the violation of Bell's inequalities should be of our least concern. As noted by Clauser and his coauthors in \cite{ClauserSuperdeterminism}, the whole pursuit of science would be meaningless: 
\begin{quote}
``Unless we proceed under the assumption that hidden conspiracies of this sort do not occur, we have abandoned in advance the whole enterprise of discovering the laws of nature by experimentation''.
\end{quote}
A similar conclusion is echoed by Zeilinger in \cite{ZeilingerSuperdeterminism}, where he stated that, if superdeterminism were true, ``it would make no sense at all to ask nature questions in an experiment, since then nature could determine what our questions are, and that could guide our questions such that we arrive at a false picture of nature''. Actually, there can be much worse implications of superdeterminism than making science pointless. Following the same reasoning, the Earth could be right now flooded with hostile species of reptilian aliens that are literally all around us, but so it happens that some regrettable sets of circumstances always prevent us from looking in the right direction at the right moment -- just as fine-tuned sets of circumstances dictate Alice and Bob's decisions how to turn their polarizers. And we could never know -- unless, at some moment, the script dictates that aliens begin to devour us alive.

Instead of as a serious hypothesis about nature, the superdetrminism can be {\tmpcolor{purple} rather} seen as a testament to the price some people are ready to pay, just to avoid the inevitable conclusion that our universe is not locally realistic (seemingly, some scientists are ready to entertain superdeterminism at least partially seriously \cite{BellStanford}). This enormity of the price only underlines the depth of the change in our worldviews that Bell's inequality and its violation force us to make.

\section{Epilogue}

\subsection{Nobel prizes}

Luckily, times are changing. With only a handful of citations in the first decade, as of today, Bell's 1964 paper \cite{Bell64} has amassed more than 27000 citations. Implications of its violation are widely discussed and brought into the spotlight of broader audiences. Finally, last year, even the Nobel committee decided to acknowledge the momentous scientific and philosophical contributions of Bell and the experimental physicists who tested his inequality.

John Clauser, Alain Aspect, and Anton Zeilinger shared the 2022 Nobel Prize in Physics. The only reason why Bell was not among the last year's laureates was that he unfortunately died already in 1990, at the age of 62. Reportedly, he was nominated for the Nobel prize in the year he passed away \cite{BellNobel}. Of the other main actors in this drama, Einstein and Bohr did not live to see Bell's theorem and the outcomes of experimental Bell tests, but both were awarded Nobel prizes for other achievements. Stuart Freedman, who assisted J. Clauser in his experiment, also died prematurely.

Taking into account that the first experimental test of Bell's inequality happened already in 1972 (and that hardly anyone expected any remaining loopholes to change our conclusions), it is reasonable to ask why it took an entire half of a century for the Nobel committee to effectively recognize, by its decision, the depth of the irrevocable shift in our perception of reality that quantum mechanics brought, and Bell proved. Stated differently, what prompted the Nobel committee to suddenly remember Bell nowadays, 50 years after the first test and 24 years after Zeilinger closed the locality loophole?

At least a part of the answer certainly has to do with the shift from pure science and philosophy into technology. As it often happens, the most fundamental research in physics, aimed at purely philosophical topics, eventually also produced some very practical spin-offs. At some point, it was realized that the strange non-classical properties of quantum entanglement -- the phenomenon first brought to attention in the EPR paper -- can be used in many technologies, from cryptography, highly sensitive measurement techniques, superdense information coding, all the way to the entirely novel concept of quantum computing. It actually opened a whole new branch of research called quantum information science. In this light, also the achievements of the pioneers of this field, both in the theory and experiment, were gradually more and more difficult to ignore.

\subsection{The Aftermath: Reality or Locality}

And, when the dust has settled, what is that this ``most profound discovery of science'', as Henry Stapp praised Bell's inequality and its violation, actually tells us? \cor{While this obviously is the most important question in this context, it, to this day, remains a controversial one.}

On multiple occasions we have stated that Bell's violations falsify the idea of ``local realism'' -- it shows us that nature certainly cannot be both ``local'' and ``real'' at the same time. While it might be neither. This common formulation of conclusions dates back to Clauser and Shimony's 1978 paper \cite{ClauserShimony1978}. While the idea of the locality they had in mind is fairly clear (i.e.\ that no influence can be exerted immediately to arbitrary distance, i.e.\ as if mediated at infinite speed), the notion of realism they explain as follows: ``Realism is a philosophical view in which external reality is assumed to exist and have definite properties, whether or not they are observed by someone.'' Giving up the notion of objective reality independent of observers, which was the cornerstone of exact science from its advent, might seem like a quite steep price to pay. The problem with violating the notion of locality, on the other hand, is not only in the fact that it is not in the spirit of Einstenin's theory of relativity (it neither strictly experimentally contradicts it, since particle entanglement does not allow sending information faster than light). The additional problem is that giving up locality also jeopardizes the idea that nature can be split into separate subsystems and investigated in this way. If whatever happens in our lab can be, in principle, influenced by what is at the moment happening in another galaxy, that undermines the very reductionist approach to reality. It certainly does away with the good old mechanist ideal of reality being a set of moving pieces that influence only those in the immediate vicinity (the intuitive idea of gears and levers, or of particles bouncing off each other). We should clarify that, although the influence by ``pulling and pushing'', i.e.\ by the contact-interaction of impenetrable solid bodies was the golden standard, or at least ideal, of Newtonian classical physics, it is true that Newton himself was forced to introduce a long-range force to explain gravity. However, that idea was immediately considered as a sort of occult and definitely troubling, even by Newton \cite{BohmMetaphysics}. And what was merely shunned in Newton's era, would eventually become inadmissible with Einstein's theory of relativity due to the light-speed limit on all influences. It took a while for science to get rid of such unwelcome ``occult'' remnants: first, thanks to Maxwell's theory of electromagnetism, which solved this problem for the case of electric and magnetic interactions, and finally by Einstein who solved the same problem for gravity. They provided perfectly local (and of course, realist) explanations of the universe's inner workings. With these achievements, the Newtonian ideal of a mechanistic universe was seemingly completely accomplished and the universe fully gained the intuitive feel of a huge clockwork mechanism. Yet, the option opened by Bell's test -- to abandon the locality of physics, all of sudden endangered this ancient worldview, in a way that was far more disturbing than was the simple inverse-square Newton's law of gravity.

Even without these purely philosophical repercussions, the huge problem of internal compatibility between nonlocal interactions and the theory of relativity means that accepting the nonlocality requires serious revisions of our understanding of space and time. Overall, as Clauser and Shimony emphasize in their paper \cite{ClauserShimony1978}:
\begin{quote} 
``The conclusions [from Bell's Theorem] are philosophically startling; either one must totally abandon the realistic philosophy of most working scientists, or dramatically revise our concept of space-time''.
\end{quote}

{\tmpcolor{purple}We stress again that the issue of determinism is unrelated here. First we must clarify that, per se, realism does not inevitably entail determinism (though the two notions are often conflated). It is easy to conceive a model that would be realistic (i.e.\ systems have definite properties irrespective of measurement/observation) and local in spite of being stochastic (i.e.\ laws of physics do not allow us to predict actual outcomes of experiments, but only probabilities of different outcomes). For example, in the Bell's context, we might consider a hypothetical model in which photons always do have well defined polarizations (hence the model is realistic), while it is nevertheless impossible to surely predict whether a particular photon will pass the polarizing filter or not (thus model is indeterministic): in particular, the probability of transmission could be proportional to the cosine squared of the angle between photon polarization and the filter orientation (such dynamics would be local and would correctly reproduce the experimentally observed Malus law).\footnote{\cor{At first sight, it might seem possible to trade this randomness for a complex but again deterministic law of nature,} if we would allow each photon to somehow possess within itself a long enough predefined sequence of random numbers: each time it would encounter a polarizing filter, it would then ``consult'' this list and ``decide'' according to some deterministic algorithm whether to pass or not, instead of this outcome being truly undetermined until that moment. However, if each photon possessed a different list, this would reveal itself via violation of Bose-Einstein statistics, as they would cease to be identical particles. And if this list would be identical for all photons, then it would effectively constitute a new deterministic law of physics -- something that should be expected to reveal itself via repeatable patterns in behavior of particles under identical conditions. Finally, if every particle could somehow ``consult'' random-like number list that is external to the particle (and thus not violate statistics of identical particles), this would be problematic from the locality perspective. This last option includes also the so-called ``second time around'' universe: assume there is a nondeterministic universe, we let it evolve and ``record'' everything that happens; if we then "play" this recorded universe, all events would be predetermined and thus the universe deterministic, but its dynamical laws would not be local. Thus, non-deterministic locally-realistic models arguably constitute an entirely separate class.}
		
Importantly, giving up determinism cannot save the role of local realism in nature -- since locally-realistic models like the one above must also satisfy Bell's inequalities, in spite that they have sacrificed determinism. This fact was obscured in the original formulation of Bell's inequality (which actually presumed determinism), but was recognized in later generalizations of the theorem.}

\subsection{Locality?}

On the other hand, some of the more recent discussions on the subject tend to question this Clauser's ``reality or locality'' formulation of the conclusions \cite{BellStanford}. \cor{It is often argued that Bell's inequality violation implies some elements of nonlocality irrespectively of our stance on realism, i.e.\ that ``locality'' assumption cannot be saved -- at least not in its standard sense -- even if we give up realism. Emphasizing these nonlocality features,} some modern scholars thus prefer to summarize implications of Bell's violation simply as a constraint that any hidden-variable theory must be nonlocal, more in line with Bell's own formulation of his theorem: ``If [a hidden-variable theory] is local it will not agree with quantum mechanics, and if it agrees with quantum mechanics it will not be local'' \cite{BellSpeakable}. \cor{To understand this emphasis on the hidden variables, but also to better understand the idea of sacrificing locality to preserve realism, we must return again to Bohm-de Broigle's hidden-variable interpretation, aka pilot-wave theory (PWT) \cite{BohmianMechanicsStanford, BohmianMechanics}, which hugely influenced Bell.

Namely, this theory was, and still is, basically the only known model of nature that explicitly does away with locality premise in order to retain realism in a sense somewhat close to the classical one. (Though, even there the concept of reality had to be extended to include a quite abstract wavefunction that lives in a mathematically involved construct of a so-called ``tensor product of single-particle Hilbert spaces''}, instead of keeping only the intuitively clear picture of the classical three-dimensional space.) Bohmian mechanics (as PWT interpretation of quantum mechanics is often called) assumes the existence of true and precise positions of all particles -- here playing the role of hidden variables -- which are well defined at all times, irrespectively of observation (in spite that knowledge about these positions, according to PWT, is not experimentally accessible). Besides, the theory is deterministic. However, \cor{as we already discussed,} the price is paid in the form of manifest nonlocality: the velocity of every particle depends on the current positions of all particles that exist in the universe, in a way determined by an additional ``guiding'' equation. This remote influence, which exists irrespective of the mutual distance between particles, is non-local simply by the definition and not as a consequence of any interaction potential. Apart from the guiding equation, this theory also postulates Schrodinger's equation that governs the evolution of the wavefunction in precisely the same way as in standard quantum mechanics. The theory is constructed in such a way as to produce experimental predictions mathematically equivalent to predictions of the standard quantum formalism.\footnote{\cor{After assuming initial probability densities that match $|\psi|^2$.}} This is possible since those hypothetical real positions of particles (i.e.\ hidden variables) do not affect the dynamics in any way -- the latter is entirely determined by the standard quantum mechanical wavefunction. Since these positions can be neither inferred by measurement, their main purpose is to provide philosophical comfort of allowing one to say that the particles actually had well-defined properties (more precisely, well-defined positions -- things become much more complicated with other properties like spin), in spite of our eternal lack of precise knowledge about them. Of course, it depends upon personal philosophical predilections to whom will such a possibility sound comforting, and to whom misleading and strikingly \cor{non-parsimonious}. In fact, the facets of this interpretation that its proponents find most appealing are often the same ones that the critics of the theory see as unnecessary and extravagant additions to the complexity of an already complete theory. Bohmian mechanics is often criticized from the viewpoint of Occam's razor, as it requires the introduction of additional properties and an entirely new dynamical equation, \cor{and} it also must go to lengths to solve technical issues nonexistent in the standard formulation \cor{--} while at the same time it is not adding anything to the explanatory power of quantum mechanics.

But, philosophical preferences aside (which can hugely differ, as in the case of Einstein and Bohr), the inherent non-locality of PWT inevitably leads to both conceptual and severe technical problems in attempts to extend nonrelativistic quantum mechanics into the relativistic domain, and especially to the deeper, underlying quantum field theory \cite{BohmianMechanicsStanford}. This was likely the main reason why Einstein was not satisfied with Bohm's proposal, in spite of its realism (besides, he found it ``too cheap'' \cite{BellCheapBohm}). Bell, \cor{while favoring this view of quantum mechanics, was also aware} of its problem with non-locality. In his \cor{opinion, the PWT} would have been a perfect theory, if it was not for this problem. As we explained, Bell eventually arrived at his theorem by attempting to answer the question of whether it was possible to improve the Bohmian mechanics in a way that would retain its realism but also remove the troubling non-local features (which motivated his formulation of the theorem statement via hidden variables). The conclusion of Bell's theorem -- that no such improved hidden-variable theory can exist that would be compatible with predictions of quantum mechanics -- was immediately disappointing to the author since the prospect of experimentally invalidating quantum theory sounded implausible. Once the Bell tests have indeed confirmed the validity of quantum predictions, Bell's theorem from then on implied that any hidden-variable theory, which aspires to conform to the experiments, must suffer from the same non-locality problem as the Bohmian mechanics. 

\cor{But} instead of being viewed as a deal-breaker for hidden variable theories in general, by proponents of PWT, this was seen as an argument in favor of their interpretation: the non-locality was no longer the fault of the particular Bohmian approach, because now it was clear that no other hidden variable theory can do better, even in principle. Moreover, advocates of Bohmian mechanics particularly tend to emphasize that some sort of nonlocality is unavoidably inherent in quantum theory, per se. \cor{After all, as pointed already in the EPR paper, even the strict formalism of quantum mechanics, in the form advocated by ``anti-realist'' Bohr, predicts that Alice can (to some extent) nonlocally influence the wavefunction of Bob's particle -- in whatever way we interpret wavefunction, this can be seen as some sort of nonlocality.\footnote{Of course, much depends on the precise mathematical definition of ``locality'', which varies in the literature.}} At the same time, they mostly dismiss that Bell's inequality violations should have any logical implications on the idea of realism (even the mention of the existence of “hidden variables” – which is certainly an assumption about reality – is often avoided). For example, for Tim Maudlin, who is a vocal supporter of Bohmian mechanics, the implications of Bell tests can be simply summarized as ``actual physics is non-local'', period \cite{MaudlinBell}.

\subsection{Reality?}

While \cor{EPR paradox and} Bell's theorem indeed suggest that quantum mechanics has some sort of non-local elements, we believe that such formulations that ignore mentions of realism (put forward mostly by Bohmian sympathizers) are at least misleading. Formulations of this type tend to wrongly reduce the deep and universal philosophical implications of Bell's inequality violations to an ostensible dichotomy between local and non-local hidden variables. They completely ignore (or obscure) the fact that violations of Bell's inequality come as the final confirmation of Bohr's predictions, in contrast to Einstein's expectations, and that, as such, the observed violations also strongly reinforce and support Bohr's view. And this view is, in the context of EPR and Bell, succinctly formulated by Clauser in \cite{ClauserReview}:
\begin{quote}
Bohr's argument is more readily understood, once that one recognizes that it is
based on his denial of realism. That is, it is impossible to physically disturb something that doesn't exist! Thus, if there is no physical-space description of either the quantum mechanically entangled individual systems or of the actual disturbance process, then the required associated existence of non-local action-at-a-distance is a non-issue. The objectionable aspect of Bohr's description (action-at-a-distance) is then, consistently with Bohr's assertion of completeness, simply nonexistent, by definition!
\end{quote}
Accordingly, in Bohr's view, even if there is some sort of non-locality after all in quantum mechanics, it is of absolutely secondary and interpretational character. \cor{As pointed by Clauser above,} what does it mean to physically disturb something at a distance, if that something does not truly exist, in the first place? 

On the other hand, the premise of realism was built in from the outset: the ``elements of reality'' that were presumed to exist in the EPR paradox were reflected via the variable (or set of variables) $\lambda$ appearing in the Bell's proof and representing the hidden \cor{variables.} While, in the first version of Bell's theorem, these elements of reality were seen as giving rise to deterministic outcomes of the experiments, even the later versions of the theorem, which did not incorporate determinism, nevertheless relied on realism presumption. \cor{The hidden variables, which remained present in all versions of the proof, were effectively assumed to exist and have well defined values irrespective of any measurement -- reflecting the views of realists, as opposed to Bohr's understanding of physical properties.} As argued in \cite{BruknerReality}, this was an indispensable part of all proofs of the theorem since the way the joint probabilities were computed implied that the ``values coexist together independently of which experiment is actually performed on either side, and in this sense are `real'.''

A part of the problem with emphasizing non-locality at the expense of considering issues with reality might be in the suggestively chosen terminology. In usual physics jargon, non-locality commonly implies, or suggests, some sort of ``non-local interaction'', where ``interaction'' generally denotes some well-defined force, or potential\footnote{or a term in Lagrangian} that causes/mediates influence between well-defined objects of the theory (e.g.\ particles or fields). In other words, the very term ``non-locality'' is not ``ontologically neutral'', in the sense that it effectively presupposes, or at least suggests, some sort of realism (e.g.\ in the form of hidden variables involved in these interactions). Therefore, if the conclusion of Bell's violations is indeed summarized simply as ``actual physics is non-local'', that even leaves an impression that Bell's tests reinforced Einstein's position of realism, while in reality, the logical implications of Bell's violations were certainly the opposite (without doubt, realism was far more plausible hypothesis while we still thought that such theories could be constructed in full observance of the relativistic principles). Similar is true also for the formulations closer to the original Bell's statement, like ``any hidden variable theory must be non-local'' -- again connoting that the entire issue is a dichotomy between local and non-local hidden variable theories, instead of between the realist Einstain's viewpoint and the ``non-realist'' Bohr's stance.

A much better choice would be the word ``inseparable'' (this word is also more closely related to the mathematical term ``non-factorizable'', which better corresponds to the presumptions of Bell's theorem than ``non-locality'' \cite{BellStanford}). For, if the conclusion is that physical systems on Alice and Bob's side cannot be thought of as separate\footnote{\cor{There is also a formal definition of \emph{separability} in quantum information theory (being the opposite of entanglement), that well corresponds to this intuitive meaning.}} entities, then it is a statement that pertains to our understanding of reality as a whole (suggesting it possesses a certain holistic character) and not merely to some technical details about the speeds at which the influences can propagate. 

Besides, if the essence was truly in non-local interactions that somehow allow particles to ``communicate'' faster than light, we could naturally expect that also we -- and not only particles among themselves -- should be able to send faster-than-light signals. On the contrary, this impossibility to “operationally” violate the locality principle (i.e., by sending faster-than-light signals), actually shows that any definition of “locality” – according to which quantum mechanics can be non-local – must heavily rely on certain assumptions about reality. Obviously, if we constrain the definition of reality to only observably measurable effects (in the extreme, to our perceptions of measurement outcomes), we are led to conclude that nature is absolutely local: due to no-signaling constraint, nothing Alice does will ever immediately influence Bob's faraway observable experiment outcomes (or his perceptions of these outcomes). Therefore, if we are to claim the existence of any sort of non-locality, we must first extend this ``operationalist'' definition of reality, to include elements not directly observable but nevertheless ``real''. \cor{An obvious way is to assume the existence of hidden variables, which was done in Bell's theorem and led to the violation of locality. Another is to take a realist stance on the wavefunction (as opposed to epistemic): if the wavefunction is taken to be literally real, then the influence of Alice's measurement on the distant wavefunction of Bob's particle, noted in the EPR paper, already demonstrates the non-locality of nature. (We stress that it is not necessary to see the wavefunction as really existing in physical space, since is not directly observable -- formally, it is a mathematical tool used to compute probabilities.) In any case, the fact that we cannot even speak about non-locality here without previously introducing some nontrivial premises about reality shows that what is truly violated is rather a} combination of locality and certain intuitive assumptions about reality, and such combination is maybe better captured by the word ``separability'' \cite{BohmMetaphysics}.

As noted in \cite{arxiv}, in his reply to the EPR paradox Bohr precisely argued that the notion of separability, as understood by classical physics and Einstein, is no longer applicable to the reality of quantum mechanics. In \cite{EPRreply} Bohr writes ``...we have to do with a feature of {\it individuality} completely foreign to classical physics'', putting the word ``individuality'' in italics. Seemingly, his point was that quantum mechanics does not allow us to fully consider subsystems individually, as we were used to in classical thinking, since reality is in some sense not divisible (''individual'' comes from Latin, literally meaning ``not divisible''). So, Bohr did recognize that quantum mechanics implies some non-local effects in the sense of inseparability, but this was to him merely one of the features of that different sort of reality that we need to embrace. And this new reality clearly does not conform to Einstein's idea of classical ``realism''.

This difference between notions of locality and separability can be a bit better explained if we switch to the description using the so-called density matrix, instead of the wavefunction. The density matrix is a somewhat richer mathematical object than the wavefunction, capable of jointly describing both the quantum-mechanical lack of knowledge about systems (described by superposition) and the classical lack of knowledge (described by so-called mixed states). Instead of using state-vector \cor{$\ket{\Psi}$} given by (\ref{EnSpinZ}) to describe EPR entangled state of two particles with spin, we can use the density matrix $\rho_{AB}$ that corresponds to this \cor{state} and describes the joint system of both particles (where $A$ in the subscript stands for Alice, and $B$ for Bob).\footnote{\cor{It is a two-particle operator defined as $\rho_{AB} = \ket{\Psi}\bra{\Psi}$.}} However, in addition to the density matrix that describes the joint system, we can define also density matrices corresponding to each of the particles taken individually.\footnote{\cor{They are obtained from $\rho_{AB}$ by taking the ``partial trace'' over the other space, e.g.: $\rho_A = \mathrm{tr}_B( \rho_{AB})$.  The result is a single-particle operator.}} These mathematical objects deal with only one of the particles and do not have anything to do with the other one. As if we took the joint two-particle description and then forgotten (erased) anything about the second particle and their mutual connection. It turns out that such density matrices of individual particles (subsystems) provide sufficient information to predict the outcome \emph{of any possible} experiment carried out on that particle alone! In other words, it can be argued that this reduced density matrix of the single particle is what truly describes the ``reality'' of that particle. (The analog is not possible to define on the level of wavefunctions.) Importantly, it turns out that nothing that Alice does can influence the reality of Bob's particle defined in this sense, since the quantum formalism directly implies that Alice's actions cannot change the reduced density matrix that Bob uses to describe his particle (and to successfully predict outcome probability of any possible measurement on it). For this reason, considering reduced density matrices is mathematically the simplest way to prove that it is impossible for Alice to send any information to Bob by using entanglement -- the result that is far from obvious when considering the wavefunction.

The weirdness of quantum mechanics here appears through the fact that the reality of Alice's particle alone, combined with the reality of Bob's particle alone, does not represent the full reality of the system consisting of Alice and Bob's particles taken together. Namely, the outcomes of measurements that require access to both particles, in general, cannot be inferred (not even probabilistically) based only on these two reduced density matrices, \cor{in a way that has no classical analogs.\footnote{In classical physics, a superficially analogous situation is possible only when we are lacking some piece of information. For example, we could have pairs of particles where Alice's spin is pointing upwards, and Bob's spin points down, and pairs where, vice-versa, Alice's spin is down and Bob's is up. If, from a mixture of ten pairs of the first type and ten pairs of the second type we would be randomly given a single pair (so that \emph{we do not know} from which set it originates), our (classical) description of such a pair would be superficially similar to the quantum state of entangled spins (\ref{EnSpinZ}) discussed in the context of the EPR paradox. Namely, it would also have the properties that probability of both Alice's and Bob's particle spin to be up is 50 \% (same for down), while each time Alice's spin turns out to be up, Bob's will certainly be down, and vice-versa. However, such classical correlations are always a consequence of some type of our ignorance (here, we simply did not know from which of the two sets our pair originated), but are never the feature of the system \emph{per se} -- since, classically, particles do have well-defined properties and must ``themselves know from which set they originated''. Besides, such correlations cannot violate Bell's inequality. The essence of the EPR paradox and Bell's violation can be seen from this perspective: Einstein assumed that the only possible cause of such correlations in nature must be in our ignorance, i.e.\ that there must remain some hidden variables which make quantum mechanical description incomplete. But the violation of Bell's inequality showed us that quantum correlations are not always of this ``ignorance'' type.} The simplest example of such a joint-type experiment is measuring the spins of both particles and \emph{comparing} the results. To predict outcomes of such measurements (in general) we need the full density matrix of the joint system -- and this matrix therefore somehow represents the reality of the two particles taken as a whole. As the joint matrix generally cannot be reconstructed from the reduced density matrices\footnote{And, importantly for the argument, according to quantum mechanics this is not a consequence of our incomplete knowledge, as would be in the analogous case of correlated systems in classical physics.}, the full reality turns out to be more than the sum of its parts.}

In the case of the two EPR particles in state (\ref{EnSpinZ}) the spin state of each particle is unknown, and thus undefined (the reduced density matrices are proportional to unit matrices). In the language of Bohr, individual spin components of particles do not exist. However, in spite of the non-existence (indefiniteness) of individual spins, the fact that the spins are opposite is well-defined and thus very much existent (otherwise, conservation of angular momentum and rotational symmetry would be violated). As if the two Bertlman's socks were of different colors in spite of that the color of each individual sock is being undefined/non-existent. And the reality of this ``oppositeness'' (encoded in the joint density matrix $\rho_{AB}$), is not contained in the realities of the two particles taken separately (given by the two reduced density matrices) -- defying, in a sense, the very reductionist approach to nature.

This view is intuitively easier to grasp from the viewpoint of interpretations that are closer to Wheeler's ``it from bit'' doctrine, that is those in which, in one way or the other, information plays the fundamental (ontological) role. Wheeler used to liken the universe with the ``twenty questions'' game\footnote{A spoken parlor game, where ``questioners'' try to figure out the object imagined by the ``answerer'', by posing yes-or-no questions.} where the object we are trying to guess simply was not yet decided by the answerer \cite{Wheeler20Questions}: our questions posed to nature by experiments and observations do not \emph{find something out} about the fixed reality that is out there -- instead, our questions gradually \emph{shape} that reality. The answer is, in a certain sense, \emph{created} only after the question, and we are only granted to always obtain mutually consistent answers. Returning to the case at hand, from this viewpoint there is nothing surprising in the fact that the two spins can be opposite and this fact real, as this oppositeness follows from available information, while the individual spins, about which we can know nothing even in principle, are not yet defined. By our measurements, required to prepare the initial state of the particles, we obtained confirmation from nature that the spins are opposite and that is all there is -- before we ask anything about the individual spins, nature does not have to yet decide anything about them, and they remain vague, undefined.

However, if we follow this line of reasoning, we should be also careful about who precisely is posing the questions to nature. For example, in the EPR case, not only that we have two subsystems, but also two observers, so the density matrices used to describe reality by Alice and Bob can easily start to differ. This is a consequence of the fact that the density matrices take into account our classical knowledge (and the lack thereof). Namely, when Alice obtains the result of measurement of spin projection on her particle, this new information pertains also to the joint system of the two particles. She may also \emph{update her knowledge} about the reality of Bob's particle spin (since the spins are opposite) -- and the latter now also becomes well defined \emph{from her viewpoint}. Bob, on the other hand, does not know the outcome obtained by Alice, so he describes the joint system and the subsystems with different density matrices.\footnote{\cor{All this directly follows from the standard formalism of quantum mechanics. From the perspective of Bob, who does not know the outcome, Alice performing the measurement corresponding to hermitian operator $A$ with spectral decomposition $A = \sum_i a_i P_i$ has no effect on Bob's reduced density matrix since $\rho_B' = \mathrm{tr}_A \big(\sum_i (P_i \otimes 1) \rho_{AB} (P_i \otimes 1) \big) = \mathrm{tr}_A \big((\sum_i P_i \otimes 1) \rho_{AB}\big ) = \mathrm{tr}_A (\rho_{AB}) = \rho_B$, where we used cyclic property of trace. The only change Bob's reduced density operator undergoes in this view is of taxonomic nature: after the measurement, it becomes a mixed state of the first kind -- reflecting the classical lack of knowledge of the result that Alice obtained, while prior it was a mixed state of the second kind -- i.e.\ reflecting that it was entangled with Alice's particle. This classification of course neither affects the form of the matrix, nor the underlying reality. In interpretations of quantum mechanics in which collapse is only subjective and Alice's measurement does not result in collapse from Bob's perspective, his particle remains in the mixed state of the second kind -- reflecting the entanglement which is now not only with Alice's particle, but also with Alice's measurement apparatus/environment and Alice herself. On the other hand, for Alice who knows which specific outcome $a_i$ has occurred, the density operator of the joint system changes according to the following formula: $\rho_{AB}' = (P_i \otimes 1) \rho_{AB} (P_i \otimes 1)/\mathrm{tr}_{AB}(\rho_{AB}P_i)$, leading to a nontrivial change in the reduced density matrix of Bob's particle.}}

So, what does it mean if Alice and Bob use different density matrices to describe reality? If these matrices correspond merely to Alice's and Bob's knowledge, then this sounds quite natural, since they simply possess different sets of information about the systems they want to describe. But, if we want to assume that this knowledge also directly corresponds to a given, well-defined, unique and objective reality of the particles they describe, obviously not both descriptions can be correct. We may be tempted to say that initially, both Alice's and Bob's descriptions were the same and correctly represented the reality of Bob's particle, but after Alice's measurement she has learned a piece of information that Bob does not have, and thus her description is now correct and corresponds to true reality, while Bob's does not. If all this is so, then we indeed must concede that Alice's measurement has somehow changed the objective state of Bob's distant particle, i.e.\ that we have some sort of non-local influence.

Yet, assuming existence of objective physical reality is not necessarily in the spirit of quantum mechanics and certainly is not the only option. If we restrain from such a ``realist'' urge, we may say that there is one description (density matrix, or a vector state) and the corresponding reality related to Alice, and another one related to Bob. Taking such a view of ``two different realities'' literally to a certain extent is the position of Rovelli's ``relational interpretation'' of quantum mechanics. According to this interpretation, there is nothing anthropocentric here, i.e.\ Alice and Bob have no unique role, and we could speak about the quantum state (and ``reality'' in some sense) from the perspective of any physical system -- since the state (wavefunction, or density matrix) is here observer-dependent and just encapsulates relations between the observed system and the reference system (i.e.\ observer). This is understood in a similar way as the lengths and time intervals are relative to observers (i.e.\ reference frames) in the theory of relativity. Another way to interpret different descriptions of reality by Alice and Bob -- which may be ostensibly conflicting and yet both correct -- is offered by the ``quantum Bayesianism'' (QBist) interpretation of quantum mechanics. Here, the quantum state merely represents (rational) degrees of belief that an agent can have about the outcomes of possible measurements. Yet another would be a reading of the Copenhagen interpretation where the collapse occurs strictly subjectively, and which goes hand in hand with the ontology of idealism \cite{ToTheRescue}. In this view, the subjective experiences are the only well-defined and real things in the universe, but interpreting \emph{when} these experiences occur is highly observer-dependent. For example, Alice's subjective experience of seeing her spin measurement outcome has not occurred nor have become well-defined from Bob's viewpoint until either he  learns the result from her or performs his own spin measurement that indirectly reveals Alice's outcome.

It is beyond the scope of this paper to discuss in more detail the mentioned views. Nevertheless, it is very dubious whether, and in what sense, one can say that the universe is non-local, according to these interpretations. In all of these views, the wave-function collapse after Alice's measurement happens only from her perspective, and it is essentially nothing else but the update of knowledge (or degrees of belief) Alice has both about her and about Bob's particle. The fact that Bob's particle is distant makes the update of knowledge, per se, nothing more mysterious than if Bertlman would send a sock from his pair to Andromeda galaxy, and only afterward would present us the other one from the pair: our update of the knowledge about the distant sock based on the one that we have just witnessed would be instantaneous, irrespectively of the light years of distance. Nonetheless, we most certainly would not describe the situation as a case of faster-than-light interaction -- instead, we would simply say that we found out something about the distant sock. Note that no change in the ``reality of Bob's particle'' happens from Bob's viewpoint after Alice's measurement, and thus there is no objective non-local influence of Alice on the distant particle.

The mysterious part of these interpretations is not in their attitude towards locality, but towards reality. It is in the claim that the knowledge and information (given by state vector/density matrix) do not correspond to any observer-independent (system-independent) fixed, objective physical reality -- as such does not exist. And since this idea of objective reality is now inherited by this information-based reality, \emph{finding out} something about a distant system (which consequently also defines that revealed property) might be seen as a sort of action at a distance. However, in our view, calling this simply a ``non-local interaction'' (or a demonstration of the non-locality of the universe) is at least misleading and a stretch of definition.

There are some additional arguments that it is our understanding of reality which we must radically revise as the consequence of Bell's violations (and not merely the idea of locality). These are provided by the Kochen-Specker theorem \cite{KochenSpecker} and more recent violations of so-called Leggett (and Leggett-Garg) inequalities \cite{LeggetTheory, LeggetGargTheory}. In a similar fashion as with Bell's inequalities, these theorems and experimentally violated inequalities rule out a large class of intuitive realistic theories (e.g. macrorealistic physical theories), this time irrespectively of whether our universe respects locality or not. The corresponding experiments were again first performed by Zeilinger and his team \cite{LeggetExperiment}, and they concluded: ``giving up the concept of locality is not sufficient to be consistent with quantum experiments, unless certain intuitive features of realism are abandoned.''

While it is true that the de Broglie-Bohm hypothesis can not be falsified in this way (since it was constructed to be experimentally indistinguishable from the standard quantum mechanics), these constraints on realistic theories explain why even PWT is very far from the realism standards set by classical physics. In particular, as a general consequence of the Kochen-Specker theorem, it is not possible to construct any hidden variable model (in agreement with experiments) that would have simultaneously well-defined values of all measurable properties of systems (hidden variables must be ``contextual'', i.e.\ dependent upon the measurement ``context''). For example, in PWT interpretation, a special role is given to the position of particles and they are real in the standard, classical case. However, this is not the case with spin, which cannot be taken as a genuine/intrinsic property of particle alone, but it has to do with both the wavefunction and the current experimental setup \cite{PWTSpin}. Even Bohm himself saw this contextuality as a huge departure from the basic principles of classical physics \cite{BohmianMechanicsStanford, BohmContextuality}. Besides, while postulating a wave-function existing in the real three-dimensional space would be absolutely in the spirit of ``classical realism'', it is quite different to posit, forced by the quantum phenomenon of entanglement, a wave-function living in some abstract N-fold tensorial product of three-dimensional spaces (where N is the number of particles in the universe). Arguably, such ontology loses touch with the highest ideals of a true realist theory, where reality should be intuitively clear and a sort of palpable. (The matters are only made worse by the fact that this richness of the wave-function structure, which can never be directly observed, leads to endless ``zombie worlds'' of so-called empty branches \cite{PWTEmptyBranches}.) Thus we see that even in the pilot wave theory the core idea of realism has suffered too much to warrant claims that ramifications of Bell's inequalities can be constrained only to the issue of locality.

\subsection{Something else?}

Overall, we have seen that even if we are ready to sacrifice realism, some features of locality (at least in the sense of separability) must be lost as well, while the attempts to solve the problem by explicitly giving up locality again land far from any epitome of realism. Hence, Clauser's original formulation, simply stating that we must abandon the idea of ``local realism'', seems to be the closest to the mark and to reasonably summarize the true philosophical implications of the violations of Bell inequalities. It is up to different interpretations of quantum mechanics to emphasize either the ``locality'' or the ``reality'' part of the statement.

However, we must mention the option that retains both the ideal of the locality of influences and the idea of observer-independent reality, at the expense of giving up a seemingly necessary and tacitly always implied assumption that experiments have unique outcomes. This is the widely known many-worlds interpretation (MWI) of quantum mechanics. It asserts that it is wrong to say that Alice in the EPR setting will obtain either a spin-up or spin-down outcome in the z-spin measurement -- instead, the universe will split (or multiply in some sense) into two universes, identical in every minute detail, apart from the fact that in one of them, there will be Alice seeing the up outcome, whereas the other will be inhabited by Alice registering the opposite outcome. While, as with every other interpretation, this one also has its strengths and weaknesses (which we will not delve into), it is not difficult to see that the proof of Bell's theorem cannot be carried out with so a wildly altered definition of measurement.

Many-world interpretation posits that the wavefunction of the universe is ontologically real, and observer-independent. In this sense, this is a realist interpretation. On the other hand, merely saying that MWI calls for a revision of our understanding of reality would be an understatement. Instead of having a single reality of three-dimensional space, we now have an almost infinite number of concurrent realities, i.e.\ ``worlds'', and essentially everything that might have happened has actually happened in (at least) one of these worlds. Obviously, this picture of reality is far remote from any stretch of Newtonian or classical-like worldview.

\section{Conclusion: the fate of the clockwork mechanism}

While there might not be a consensus on whether it is the idea of separability (locality) or reality that should give in, or maybe that we live in countless parallel universes, there is an absolute consensus in contemporary physics that the hopes of ever returning to anything resembling classical physics are long over. Between the abandoning of reality or locality, the philosophy of mechanistic materialism is caught between a rock and a hard place, but with no more room left in the middle. Needless to say, the many-world option is of no help here. Therefore, the very mechanism idea, that the world can be understood as a separable set of moving/acting parts that affect each other in the immediate vicinity, the view that was so pervasive in our science and technology of the last couple of centuries -- has to be forsaken.

Actually, even before Bell's theorem, and long, long before closing all loopholes in Bell tests, the overwhelming majority of eminent physicists involved in the development of quantum theory were already making radical shifts from the old philosophical views. Einstein, who still believed that the universe is something akin to a huge deterministic clockwork mechanism, was essentially the sole exception among this elite -- and his expectations in this context were, as we have seen, explicitly proven wrong.

On the other hand, Bohr with his ``anti-realist'' stance was neither alone nor the most radical. Unsurprisingly, his close collaborator Heisenberg also deemed the speculations that ``behind the perceived statistical world there still hides a 'real' world in which causality holds'', in his own words, ``fruitless and senseless'' \cite{HeisenbergQ1}. No doubt that in his opinion \cor{it was} the ``reality'' which we misunderstood in the previous centuries. He saw implications of quantum mechanics as a call for a switch from Democritus' to Plato's views \cite{HeisenbergQ2}: 
\begin{quote}
``I think that modern physics has definitely decided in favor of Plato. In fact the smallest units of matter are not physical objects in the ordinary sense; they are forms, ideas which can be expressed unambiguously only in mathematical language.''  
\end{quote}
Non-separability was, to him, seemingly just an anticipated part of this ideological shift \cite{HeisenbergQ3}: 
\begin{quote}
``There is a fundamental error in separating the parts from the whole, the mistake of atomizing what should not be atomized. Unity and complementarity constitute reality.''
\end{quote}

Wolfgang Pauli was also clear that it is time to move on from the worldview based on classical prejudices \cite{PauliQ1}: 
\begin{quote}
``The mechanistic world view seems to us as a historically understandable, excusable, maybe even temporarily useful, yet on the whole artificial hypothesis.'' 
\end{quote}
Fascinated by Jung’s idea of synchronicity (which he saw on par with the principle of causality, complementing it in the explanation of the world) and psychology of the unconscious, Pauli's views are generally classified under the heading of a dual-aspect theory and he has never highly esteemed the philosophy of materialism \cite{PauliQ2}.

Interestingly, while Erwin Schrodinger often supported Einstein in debates on topics in physics, his basic philosophical views were even more distant from materialism than Pauli's. Finding agreement in Upanishads, he saw Brahman and Atman as essentially the same, clearly supporting idealist philosophical views: ``The external world and consciousness are one and the same thing'' \cite{SchrodingerQ1}. He leaves no doubt which of the two is the fundamental one in his view \cite{SchrodingerQ2}:
\begin{quote}
``Consciousness cannot be accounted for in physical terms. For consciousness is absolutely fundamental. It cannot be accounted for in terms of anything else.'' 
\end{quote}
Here, by consciousness he does not consider merely individual, personal consciousness, since he finds the plurality of consciousness to be deception (Indian \emph{maya}): ``The ego or its separation is an illusion'' \cite{SchrodingerQ3}.

Idealist views were expressed already by Max Planck: ``I regard consciousness as fundamental. I regard matter as derivative from consciousness.'' \cite{PlanckQ1} In other words, his opinion was that ``mind is the ultimate source of matter.'' \cite{PlanckQ2}

Consciousness was also essential for Eugene Wigner and John von Neumann. They were even of the opinion that it is indispensable for the quantum theory in the sense that, according to Neumann–Wigner interpretation, it is the consciousness of the observer that causes the collapse of the wavefunction. Von Neumann also recognized that the wavefunction collapse (''the process 1'' as he called it) has a unique role in the universe as only it can create new information (unlike the process 2 -- the deterministic evolution by Schrodinger's equation).

Emphasis on both information and the crucial role of the observer we also encounter in the views of John Archibald Wheeler, in his already mentioned ``it from bit'' doctrine, combined with the idea of a ``participatory universe'' (Participatory Anthropic Principle). For him (conscious) observers were participators in creating the reality, by posing questions to nature (by measurement and observation), as in his variation of the ``twenty questions'' game. The information contained in the answers was the basis for the existence of all physical things.

Of all of them, the philosophical positions of Max Born were the least radical (with respect to the prevailing materialistic views of that time). Since his contribution to quantum mechanics was the probabilistic interpretation of the (modulus squared) wavefunction he, naturally, disagreed with Einstein on the issue of determinism. This further led to differences about the existence of free will, since Born opinion was that the breach of causality arising from quantum mechanics provides a channel for our true agency in the physical world \cite{BornIndeterminism}. In this sense, he also was departing from the philosophy of radical (monistic) materialism.

Curiously, even David Bohm did not support Einstein's philosophical views. This might come as a surprise, since it was he who revived and further developed de Broglie's vague idea of pilot-wave theory, to fully develop what is now known as de Broglie–Bohm theory -- and is often seen as the last refuge of people looking to salvage as much as possible of our classical intuitions about the world and objective reality. (As for de Broglie, it is known that he changed his philosophical views a few times.) But, Bohm had far more complex metaphysical intentions than attempts to resurrect mechanical worldview. To him, the pilot-wave theory was a way to unite his holistic view of the universe, with the importance of information and mind \cite{BohmMetaphysics}. He strongly believed that the reductionistic approach of classical physics, relying on the separability of subsystems, was wrong \cite{BohmQ1}:
\begin{quote}
''Ultimately, the entire universe (with all its “particles,” including those constituting human beings, their laboratories, observing instruments, etc.) has to be understood as a single undivided whole, in which analysis into separately and independently existent parts has no fundamental status.''
\end{quote}
As we can see, it turns out that Bohm advocated his mechanics not in spite of its non-locality (i.e.\ inseparability) features, but because of them. Furthermore, he saw the wavefunction as embodying a special kind of information he called ``active information'' that was per se of semantic character (unlike Shanon's information). Indeed, in this informational context, the abstract and huge dimensional space in which the \cor{wavefunction} exists has much more sense than in \cor{the context of} attempts to explain the wavefunction as a part of a classical-like realist ontology \cite{BohmMetaphysics}. And since the information was semantic, it was related to his, essentially panpsychic, view that “the particles of physics have certain primitive mind-like qualities” \cite{BohmQ2}. The main achievement of his theory, for him, was not in any underlying realism, but that ``the way could be opened for a world view in which consciousness and reality would not be fragmented from each other'' \cite{BohmQ1}. Overall, as concluded in \cite{BohmMetaphysics}: ``In the end, Bohm’s metaphysics is about as far from that of the Newtonian classical metaphysical picture of the world as one could get''. In this sense, we see that Bohm had much deeper and very different philosophical motives than most of his contemporary followers who are merely attracted by distant echoes of classical mechanistic views that can be found in Bohmian mechanics \cite{BohmMetaphysics}.

Above, we have summarized the philosophical views of the greatest physicists of the first half of the XX century, of essentially everyone who significantly contributed to the birth of the new quantum theory. (We missed mentioning Paul Dirac since he had a sort of aversion towards quantum interpretations and philosophical issues. His attitude is palpable in the following quote: ``The interpretation of quantum mechanics has been dealt with by many authors, and I do not want to discuss it here. I want to deal with more fundamental things.''\cite{DiracQ1}) Not only that Einstein was the only one of them who believed in a classical-like mechanistic conception of the world, but the vast majority of them insisted on the necessity to radically evolve our understanding of reality (not locality), often even being starkly non-materialistic and emphasizing the role of consciousness in one or the other way. And all of them were exceptional geniuses, who had the deepest available knowledge of the laws of our universe. Thus, it is safe to say that there were plenty of rational arguments that mechanical worldview was no longer a plausible description of the universe, even long before we got the rock-solid mathematical proof for that, in the form of Bell's violations.

And, finally, once it was experimentally established that Bell's inequalities are indeed violated in our universe, we encountered a truly unique situation in the entire history of science. Never before has humanity been in the necessity to abandon an entire paradigm because of a proof of mathematical nature that could \emph{guarantee} that our previous scientific view -- in a quite broad sense -- was plainly wrong. Namely, it was common, throughout the history of science, that one particular theory would succeed the previous one -- as soon as it demonstrated to provide a better and more precise description of nature. But, this time what was falsified was not some particular theory or model -- it was a whole huge class of theories, and of precisely those theories that we, just until recently, thought were the only reasonable and possible choices. We are speaking of abandoning the entire scientific worldview that was absolutely dominant for a few centuries (at least in exact sciences). It was dominant to the extent that we, for the most part, tacitly understood it surely must be the correct one, so that rarely anyone even bothered to question it. And, if it was not for Bell's theorem, which possesses this unyielding power of a precise \emph{mathematical} statement, to this day we would not be aware of the immensity and inevitability of the quantum revolution that took place. Without Bell, some physicists could be, even nowadays, seeing quantum mechanics as yet another incremental step in scientific progress, possibly \cor{one of only} technical nature. And there would always be a remaining chance that Einstein was right after all, and that someday we would find a reasonable mechanistic \cor{model,} ``rationally'' explaining all that quantum ``mumbo-jumbo'' \cor{in} a form of some good old (complicated or not) clockwork mechanism. But, thanks to Bell's theorem we today know this can \emph{never} happen. (Again, we stress that by a clockwork-like mechanistic model, we assume one being both local and realistic.) The word ``never'' is something we are basically not allowed to use in physics, apart from in this context -- such is the extraordinary generality and pure mathematical strength of Bell's ``no go'' theorem. This is the reason, and a much of justification, why Stapp saw Bell's theorem as the most profound in all of science.

With such few things that we can declare with certainty in physics, and with the debunking of mechanistic doctrine after Bell being one rare example, it is unbelievable that the perception of the universe as a (deterministic) clockwork mechanism persists to this day, even among many (obviously less well informed) scholars of various scientific disciplines, and let alone in the general public. 
Ironically, most of them are clinging to mechanistic materialism in a false belief that they are upholding strict scientific views. We hope that the Nobel prizes awarded last year can help raise awareness of the scientific and philosophical revolution that happened almost half a century ago (with the first Bell tests) and that a clockwork mechanism picture of reality will soon be widely recognized for what it is -- a ``flat Earth'' of philosophy.

\section*{Acknowledgment}

This research was supported by the Science Fund of the Republic of Serbia, grant 7745968, ``Quantum Gravity from Higher Gauge Theory 2021'' --- QGHG-2021. The contents of this publication are the sole responsibility of the authors and can in no way be taken to reflect the views of the Science Fund of the Republic of Serbia.

\appendix

\section{Bell's inequality}
\label{app:inequality}

A complete understanding of how it was possible to formulate and prove a statement so general as that of Bell's inequality can be gained only by going through its derivation. Thus we will briefly derive the simplest version of Bell's inequalities known by the name CHSH inequality (in years after Bell's original paper \cite{Bell64}, many variations of Bell's initial inequality appeared, all of them nowadays collectively called Bell's type inequalities, or simply, Bell's inequalities; CHSH was named after John Clauser, Michael Horne, Abner Shimony, and Richard Holt).

\cor{For simplicity of exposition of the proof, instead of considering the state of two particles with spins $\frac 12$, we will here analyze} a system made of two photons and consider their polarizations. \cor{(Beside mathematical convenience, most of the real experiments were performed with photons, so this different choice of physical system also simplifies comparisons with actual tests of Bell inequality)}. We will \cor{again assume} that the first photon is heading towards Alice and her lab, while the second is heading towards Bob and his lab, which we will take to be far apart. While we were previously considering two particles with \cor{opposite spin directions, now we will focus on a case where the two photons are emitted (in opposite directions) from a common source in such a way that they have identical polarizations. Throughout this section we will keep in mind a realistic picture of the universe, where the polarizations are existing and are well-defined irrespective of observation -- and thus we will assume both photons have the same, real polarization angle, to be denoted as $\lambda$. (As} already stressed, the inequality we are about to prove puts a constraint on locally-realistic universes). This value $\lambda$ is unbeknown both to Alice and Bob, and might be thought of (in the simplest case) as a random angle (not necessarily in the sense of fundamentally-inherently random).\footnote{\cor{Eventually, will turn out that these assumptions of identical polarizations $\lambda$ of both photons and of random distributions are not essential for the derivation of CHSH inequality.}}  

Quite generally, if we direct an incoming photon towards a polarizing filter, it will either pass the filter or be absorbed. In principle, we expect this outcome to depend at least on two factors: the angle of polarization of the photon, and the angle at which we orient our polarizing filter. (Commonly, it is understood that if the two angles coincide, the photon will certainly pass the filter, while if the angles are perpendicular the photon cannot pass. However, for this derivation we will not take anything for granted and, in any case, this will be irrelevant for the derivation.)

Both Alice and Bob will let their photons impinge on polarization filters: Alice will orient her filter at an angle to be denoted as $a$, while Bob will orient his filter at angle $b$. We will introduce a variable $A$, which will take value 1 if Alice's photon manages to \cor{pass through} her filter, and value $-1$ if it gets absorbed. And here comes the crucial part of the entire theorem: what does the outcome, i.e.\ the $A$ value depend upon? Certainly, we expect $A$ to depend on Alice's decision at which angle to orient her polarizer (that we denoted by $a$), as well as on the angle of the photon polarization (denoted by $\lambda$). Importantly, the outcome $A$ cannot depend on Bob's choice of his angle of polarizing filter $b$, simply because Bob can be light years away from Alice, and the presumption is that the universe respects the locality principle (i.e.\ no instantaneous influences). Therefore, we may write $A$ as a function of $a$ and $\lambda$, that is $A = A(a, \lambda)$. In principle, it might be that the value $A$ also depends on some other factors -- at the end, we will see that the result trivially generalizes to include this possibility. In the same way, we introduce variable $B$ and have $B = B(b, \lambda)$.

We will consider the expected value (or mathematical expectation, similar to arithmetic mean value) of the product $A(a, \lambda) B(b, \lambda)$. Since the angle $\lambda$ is here the only unknown \cor{variable,} to find the expected value $E(A\cdot B)$ we must average over all possible angles:
\be E(A\cdot B) = \int_0^{2 \pi} A(a, \lambda) B(b, \lambda) \rho(\lambda) d\lambda. \label{EAB} \ee
Since there are infinitely many possible angles, we had to integrate over all of them to get the average. Here, $\rho(\lambda)$ denotes the probability distribution for the angle $\lambda$ -- i.e.\ probability (density) that the photons have the particular polarization angle $\lambda$ -- allowing for the possibility that some angles are more probable than the others. We only know that it must hold:
\be \int_0^{2 \pi} \rho(\lambda) d\lambda = 1, \label{rho} \ee
since it is certain that the angle $\lambda$ must take some value between $0$ and ${2 \pi}$. (If the distribution is uniform, then we simply have $\rho(\lambda) = \frac 1{2 \pi}$.)
After integration over $\lambda$, we see that the expectation value $E(A\cdot B)$ depends only on the remaining variables $a$ and $b$, that is, the averaged value of this product that we will obtain after many repeated experiments eventually depends only on the choices of angles that Alice and Bob make. For simplicity, we can thus write $E(a,b) \equiv E(A(a,\lambda) \cdot B(b, \lambda))$

Now comes the ingenious part of Bell's theorem (in the CHSH variant). We will consider a very specially defined value $S$:
\be S \equiv E(a,b) + E(a',b) + E(a,b') - E(a',b'), \label{Sdef} \ee
and shortly it will become clear why is such a choice relevant.

Above,  $a, a', b$, and $b'$ denote 4 different angles. Namely, the idea is that Alice and Bob first set their filters to angles $a$ and $b$ and find the expectation value $E(a,b)$ by averaging over many experiments. Then Alice shifts her angle to $a'$ and they find $E(a',b)$ in the same way. Next Alice sets her angle to $a$, and Bob to $b'$ to find $E(a,b')$. In the end, they also find $E(a',b')$.

Finally, we can explicitly formulate the famous Bell's inequality (in the CHSH variant): in \emph{any} locally-realistic universe, and for any choice of $a, a', b$ and $b'$, the value of $S$ must be less or equal to 2, i.e.
\be S(a, a', b, b') \leq 2. \label{inequality} \ee
Before we explain the relevance of such a result, let us first prove the inequality.

We begin by noting a simple fact: either $A(a, \lambda) + A(a', \lambda) = 0$ or $A(a, \lambda) - A(a', \lambda) = 0$. This is actually a trivial statement, since both $A(a, \lambda)$ and $A(a', \lambda)$ can take either value 1 or $-1$, so that either their sum, or their difference, must vanish. In a similarly simple manner, it follows that:
\be B(b, \lambda) (A(a, \lambda) + A(a', \lambda)) + B(b', \lambda) (A(a, \lambda) - A(a', \lambda)) \leq 2. \notag \ee
This holds since one of the two terms must vanish, and the other term cannot be greater than two by absolute value: $|A(a, \lambda) \pm A(a', \lambda)| \leq 2$ and $|B(b, \lambda)| \leq 1$, $|B(b', \lambda)| \leq 1$. Now, if we multiply this inequality by $\rho(\lambda)$ and integrate over all angles $\lambda$, we directly obtain the result we intended to prove:
\bea & \int_0^{2 \pi} \Big(B(b, \lambda) A(a, \lambda) + B(b, \lambda) A(a', \lambda) +  B(b', \lambda) A(a, \lambda) - & \notag \\ 
& B(b', \lambda) A(a', \lambda)\Big)\rho(\lambda) d\lambda \leq 2 \int_0^{2 \pi} \rho(\lambda) d\lambda. & \eea
Namely, by comparison with the definition (\ref{Sdef}) we recognize that, due to Eq. (\ref{EAB}), the integral on the left side yields precisely $S$ value, while the expression on the right-hand side \cor{is equal} to 2, due to equation (\ref{rho}).

Thus, we have proved the inequality (\ref{inequality}) which tells us something (quite specific indeed) about averaged outcomes of repeated experiments with photons and polarizers. Crucially, we obtained this result essentially without assuming anything about the physical laws that govern the polarizations and filters! The only thing we explicitly assumed was that Bob's choice of how to set up his apparatus cannot influence Alice's outcomes far, far away (and vice-versa). For simplicity, we also initially assumed that Alice's outcome depends only on her filter angle $a$ and the unknown polarization of the photon $\lambda$ (same for Bob). However, we did not need this assumption, since we can repeat the same derivation while taking $\lambda$ to represent a whole set of relevant values, instead of a single angle. Among these values in the set $\lambda$ can be, for example, the location of Alice's lab, and/or two different angles instead of one -- one for Alice's photon and a different one for Bob's, or any other value(s) that might be relevant for the outcomes. Regardless of what the set $\lambda$ contains, we will again integrate over all possibilities, weighted by the corresponding probability density $\rho(\lambda)$, to get the expected value of the product $A\cdot B$. Note that the general form of the probability density $\rho(\lambda)$ also allows for the possibility not only that some photon polarizations are more frequent than others, but also that photon polarization need not be random at all (e.g. taking $\rho(\lambda)$ proportional to delta function). It also allows for much more complex hypotheses such that polarizations on Jupiter could be more often inclined towards its moon Io, while those on Earth must be aligned to follow the angle of the Leaning Tower of Pisa. Whatever the rules are, the inequality must be satisfied!

Actually, in the derivation of (\ref{inequality}), we have tacitly used another assumption about the underlying physics: we implied it was deterministic. Namely, this is seen from our statement that Alice's outcome $A$ is a \emph{function} of values $a$ and $\lambda$, meaning that each time $a$ and $\lambda$ are the same, the outcome $A$ must also be the same. However, it turns out that the inequality remains the same even if we allow the physics to be indeterministic, and if we assume only that probabilities of getting $A=1$ or $A=-1$ depend on values $a$ and $\lambda$. (Essentially, we only need to replace the $A$ value with its mathematical expectation.)

In addition, during this entire analysis, we had in mind a picture of a well-defined objective reality -- two photons exist in and of themself, with their always well-defined properties such as polarization. And we assumed that outcomes $A$ and $B$ (or, at least, probabilities for these outcomes) depend on these real properties of physical systems (regardless of whether these properties and their values are known, and regardless of what measurements we have or have not made).

So, to summarize, we assumed a locally-realistic universe, and nothing else about the underlying physics, and obtained that the value of $S$, defined by (\ref{Sdef}), must be \emph{always} less or equal to two.

But, why is that relevant at all? This particular definition of $S$ is important since it turns out that, for certain choices of angles $a, a', b$ and $b'$, quantum mechanics predicts that the value of so-defined $S$ will be greater than 2. \cor{This we show in Appendix \ref{app:violation}.}

\section{Quantum violation}
\label{app:violation}

\cor{To show that quantum mechanics predicts violations of Bell's inequality, we will consider the following entangled state of two photons, which nowadays can be easily realized in a lab:}
\be \ket{\Phi} = \frac{1}{\sqrt 2}(\ket{H}\ket{H} + \ket{V}\ket{V}). \label{entphotons} \ee
\cor{This state} is similar to the state $\ket{\Psi}$ of two particles with opposing spins (\ref{EnSpinZ}), just that the state $\ket{\Phi}$ describes two photons whose polarizations will be the same (not opposite) upon the measurement in any chosen direction (\cor{direction of polarization is always perpendicular to the direction of photon propagation, so we here only consider these perpendicular directions}). Here, $H$ and $V$ denote, respectively, horizontal and vertical linear polarization (with respect to some, arbitrarily chosen, axes). Note, however, that individual polarizations of the two photons are not well defined in this state: when directed into a polarization filter, each of the photons has $\frac 12$ chance of passing through, irrespective of the polarizer angle. (The arbitrarily chosen horizontal and vertical directions $H$ and $V$ in expression (\ref{entphotons}) are not actually distinctive nor preferential in any sense, since this state would retain the same form when expressed in any other basis of two perpendicular directions $H'$ and $V'$ -- the state $\ket{\Phi}$ is isotropic.) In other words, the state $\ket{\Phi}$ corresponds to the situation where individual polarizations are not well-defined (so, in Bohr's sense they are not fully real), while the fact that these two ``unreal'' polarizations are mutually identical is nevertheless well-defined and real. \cor{Following the analogy with socks, here we should consider a same-color pair, and the situation would then be similar as if the colors of the two socks were not well-defined/real, but nevertheless, it was certain that the nonexisting colors are the same (so that, observing one sock automatically also determines the color of the other one).}

To show a quantum violation of the CHSH inequality, we note again that, according to quantum mechanics, the first, i.e.\ Alice's photon has a 50 percent chance to pass through Alice's filter, irrespective of the filter orientation $a$. Quantum formalism implies that, if this happens, Bob's photon polarization state will be immediately projected to the vector state that corresponds to the same angle $a$. In other words, if we have measured that Alice's photon was polarized in the direction $a$, Bob's photon sort of immediately becomes polarized in the same way. (While a detailed discussion in this direction is beyond the scope of this review, we note that the temporal order of Alice's and Bob's measurements in this settings can depend upon the velocity of the frame of reference, seemingly leading to, among some other interesting philosophical implications, strange relativity of whether an outcome was random or predetermined.) If Bob's photon next encounters Bob's polarization filter oriented at the angle $b$, quantum physics predicts that it has probability $\cos^2(a-b)$ to pass it and $\sin^2(a-b)$ to be absorbed (this also matches the classical result known as Malus law). Therefore, the probability that both photons pass their filters -- i.e.\ to have $A=1$ and $B=1$ outcomes, according to quantum mechanics, is $P(A=1, B=1) = \frac 12 \cos^2(a-b)$. Similarly, probabilities to get the other three outcomes $P(A=-1, B=1)$, $P(A=1, B=-1)$, and $P(A=-1, B=-1)$ are, respectively, $\frac 12 \sin^2(a-b)$, $\frac 12 \sin^2(a-b)$, and $\frac 12 \cos^2(a-b)$. We should notice that no ``real angle of polarization'' $\lambda$ of the two photons, or anything alike, appears in the quantum mechanical description of the situation. That is why the hypothetical angle $\lambda$ from our prior analysis of a realistic universe would be usually called a ``hidden variable''. Also, note that the probability of both photons passing their polarizers cannot be written as a product of two independent probabilities for each of the photons to pass (as we previously presumed possible and necessary in a locally-realistic universe).

Knowing these probabilities, it is now easy to compute the quantum expected value of the product $A\cdot B$:
\bea E(a,b) = E(A B) &=& P(A=1, B=1) \cdot 1 \cdot 1 \notag \\
&+& P(A=-1, B=1) \cdot (-1) \cdot 1 \notag \\
&+& P(A=1, B=-1) \cdot 1 \cdot (-1) \label{Eprob}\\
&+& P(A=-1, B=-1) \cdot (-1) \cdot (-1) \notag \\
= \cos^2(a-b) &-& \sin^2(a-b) = \cos 2(a-b). \notag \eea
Once we know the formula $E(a,b)$ for the expected value of the product for given angles $a$ and $b$, we can directly compute the $S$ value for any given choice of the four angles $a, a', b$ and $b'$. To complete Bell's proof, we should only find some concrete four angles for which the quantum prediction of the value of $S$ exceeds number 2. And, for example, the inequality is violated for the following choice: $a = 45^{\circ}, a' = 0^{\circ}, b = 22.5^{\circ} $ and $b' = 67.5^{\circ}$, because:
\bea & S = \cos (2 \times 22.5) + \cos (- 2 \times 22.5) + \cos (- 2 \times 22.5) & \notag \\
&- (\cos (-2 \times 67.5)) = 4\times \frac 1{\sqrt 2} = 2{\sqrt 2} \approx 2.828 > 2. & \notag \eea

Therefore, the standard, Bohr's version of quantum mechanics claims that Bell's inequality can be violated. In particular, that it \emph{must} be violated for the above choice of angles (it can be shown that this value of $S = 2{\sqrt 2}$ represents at the same time the maximal possible violation). But in a universe that is real in Einstein's sense and respects the locality of interactions, there is absolutely no way to obtain such a high value for $S$. Hence, the existence of this quantum mechanical example that violates Bell's inequality completes the proof of Bell's theorem -- that no local-realist theory can ever reproduce all predictions of quantum mechanics.

\cor{Consequently, we have proved that} the two worldviews (Einsten's and Bohr's) are not only philosophically different, but they can be also differentiated by an objective experiment. This was Bell's pivotal contribution to our understanding of the universe.

Bell's inequality, in principle, also provided general instructions on what should be experimentally measured to settle the dispute. One first needed to be able to create pairs of photons in the state given by (\ref{entphotons}), after which each photon from a pair should be directed to a polarizing filter oriented, respectively, at angle $45^{\circ}$ and $22.5^{\circ}$, in order to measure \cor{$E(a, b)$}. Then the experiment should be repeated with angles \cor{$0^{\circ}$ and $22.5^{\circ}$ to measure $E(a', b)$}, and so on. The particular expectation value is easily experimentally inferred (for any fixed choice of $a$ and $b$ angles) by using the first part of the equation (\ref{Eprob}), which in practice boils down to:
\be E = \frac{N_{++} - N_{+-} - N_{-+} + N_{--}}{ N_{++} + N_{+-} + N_{-+} + N_{--}}, \ee
where $N_{++}$ would be the number of events in which both photons passed their filters, $N_{+-}$ would be the number of cases that the first one passed but the second was absorbed, \cor{and so on.}

\end{document}